\documentclass[a4paper]{amsart}
\usepackage{latexsym}
\usepackage{amsmath}
\usepackage{amssymb}
\usepackage{amsfonts}
\usepackage{esint}
\usepackage{float}
\usepackage{graphicx}
\usepackage{color}
\definecolor{Blue}{rgb}{0.,0.,1.}
\definecolor{Red}{rgb}{1.,0.,0.}
\newcounter{smallarabics}
\newenvironment{arabicenumerate}
{\begin{list}{{\normalfont\textrm{(\arabic{smallarabics})}}}
  {\usecounter{smallarabics}\setlength{\itemindent}{0cm}
   \setlength{\leftmargin}{5ex}\setlength{\labelwidth}{4ex}
   \setlength{\topsep}{0.75\parsep}\setlength{\partopsep}{0ex}
   \setlength{\itemsep}{0ex}}}
{\end{list}}

\newcounter{smallroman}

\newcommand{\ben}{\begin{arabicenumerate}}  
\newcommand{\een}{\end{arabicenumerate}}

\def\init{\setcounter{equation}{0}}

\newtheorem{theoreme}{Theorem }[section]
\newtheorem{proposition}[theoreme]{Proposition}
\newtheorem{lemma}[theoreme]{Lemma}
\newtheorem{definition}[theoreme]{Definition}
\newtheorem{corollary}[theoreme]{Corollary}
\newtheorem{remark}[theoreme]{Remark}
\newtheorem{example}[theoreme]{Example}
\newcommand{\beq}{\begin{equation}}
\newcommand{\eeq}{\end{equation}}

\newcommand{\bex}{\begin{example}}
\newcommand{\eex}{\end{example}}
\def\bel{\begin{lemma}}
\def\eel{\end{lemma}}
\def\bet{\begin{theoreme}}
\def\eet{\end{theoreme}}
\def\bed{\begin{definition}}
\def\eed{\end{definition}}
\def\ber{\begin{remark}}
\def\eer{\end{remark}}

\def\rr{{\mathbb R}}
\def\zz{{\mathbb Z}}
\def\cc{{\mathbb C}}
\def\nn{{\mathbb N}}

\def\part{{\rm par}}

\def\slim{{\rm s-}\lim}
\def\wlim{{\rm w-}\lim}

\def\alg{{\rm alg}}

\def\c0inf{C_0^\infty}

\def\s{s}

\def\proof{
\noindent{\bf Proof.}\ \ }

\def\ch{{\mathfrak h}}

\def\cZ{{\mathcal Z}}

\def\cD{{\mathcal D}}

\def\cM{{\mathcal M}}

\def\CAR{{\rm CAR}}

\def\a{{\rm a}}

\def\i{{\rm i}}

\def\Dom{{\rm Dom}\,}

\def\vac{{\rm vac}}

\def\qed{$\Box$\medskip}
\def\A{{\rm A}}

\def\tz{\tilde{z}}

\def \p{ \partial}

\def\12{\frac{1}{2}}
\def\14{\frac{1}{4}}
\def\x{\langle x \rangle}
\def\xt{\frac{x}{t}}
\def\supp{{\rm supp}}

\def\e{{\rm e}}

\def\d{{\rm d}}
\def\Ran{{\rm Ran}}

\def\bbbone{{\mathchoice {\rm 1\mskip-4mu l} {\rm 1\mskip-4mu l}
{\rm 1\mskip-4.5mu l} {\rm 1\mskip-5mu l}}}
\def\one{\bbbone}

\def\ii{{\rm j}}

\def\coinf{C_0^\infty}

\def\tz{\tilde{z}}

\def \p{ \partial}

\def\12{\frac{1}{2}}
\def\x{\langle x \rangle}
\def\xt{\frac{x}{t}}
\def\supp{{\rm supp}}

\def\e{{\rm e}}

\def\Ran{{\rm Ran}}

\def\bep{\begin{proposition}}
\def\eep{\end{proposition}}

\def\s{s}

\def\CARal{{\rm C\hskip 0.25 em \hbox{\raise 1.72 ex 
\hbox{$\scriptscriptstyle\rm al$}\kern -0.57 em A}R}}

\def\otimesal{\mathop{\hbox{\raise 1.5 ex
  \hbox{$\scriptscriptstyle\rm al$}
\kern -0.92 em \hbox{$\otimes$}}}}
\def\oplusal{\mathop{\hbox{\raise 1.5 ex
  \hbox{$\scriptscriptstyle\rm al$}\kern -0.92 em \hbox{$\oplus$}}}}
\def\Gammal{\hbox{\raise 1.68 ex 
\hbox{$\scriptscriptstyle\rm al$}\kern -0.50 em $\Gamma$}}
\def\Bal{\hbox{\raise 1.68 ex 
\hbox{$\scriptscriptstyle\rm  al$}\kern -0.50 em $B$}}
\def\CARal{{\rm C\hskip 0.25 em \hbox{\raise 1.72 ex 
\hbox{$\scriptscriptstyle\rm al$}\kern -0.57 em A}R}}

\def\omvac{\omega^{V}_{0, \vac}}
\def\fA{{\mathfrak A}}
\def\tauvint{\tau^{V, {\rm int}}}
\def\tauv{\tau^{V}}
\begin{document}
\title{Hawking effect for a toy model \\ of interacting fermions}
\author{P. Bouvier}
\address{D\'epartement de Math\'ematiques, Universit\'e de Paris XI,
  91405 Orsay Cedex France} 
\email{patrick.bouvier@math.u-psud.fr}

\author{C. G\'erard}
\address{D\'epartement de Math\'ematiques, Universit\'e de Paris XI,
  91405 Orsay Cedex France} 
\email{christian.gerard@math.u-psud.fr}
\date{\today}
\begin{abstract}
 We consider  a toy model of  interacting Dirac fermions in  a  $1+1$ dimensional space time describing the exterior of a  star collapsing to a blackhole. 
 In this situation we give a rigorous proof of the {\em Hawking effect}, namely that under the associated quantum evolution, an initial vacuum state will converge when $t\to +\infty$ to a thermal state at Hawking temperature. We establish this result both for observables  falling into the blackhole along null characteristics, and for static observables.  
 We also consider the case of an interaction localized near the star boundary, obtaining similar results.
 We hence extend to an interacting model previous results of Bachelot and Melnyk, obtained for free Dirac fields.
\end{abstract}
\maketitle

\begin{center}

\end{center}
\section{Introduction}\label{sec1}\init
\subsection{Introduction}\label{sec1.1}
 The {\em Hawking effect},  see Hawking \cite{Ha}, predicts  that in a space-time describing the collapse  of a spherically symmetric star to a Schwarzschild black hole, an initial Boulware vacuum state will become an  Unruh state at the future horizon: 
 a static observer at infinity sees the Unruh state as a thermal state at Hawking temperature.

 Despite the vast physical literature on the Hawking effect, there are few mathematically rigorous justifications of the Hawking effect.
  Dimock and Kay \cite{DK1, DK2} gave a construction of the Unruh state in the Schwarzschild space-time and on its Kruskal extension, using scattering theory for Klein-Gordon fields. 
  
  The first mathematical proof of the Hawking effect, in the original setting of Hawking,  is due to Bachelot \cite{Ba0}. Bachelot  considered a linear Klein-Gordon field in the exterior of a spherically symmetric star, collapsing to a Schwarzschild black hole.   This result was extended to linear Dirac fields in the same situation, first by Bachelot \cite{Ba1}, and then by Melnyk \cite{Me}. 
  The only proof to date in a non-spherically symmetric situation is due to H\"{a}fner \cite{H}, who gave a rigorous proof of   the Hawking effect for Dirac fields for a star collapsing to a Kerr black hole.

The common theme of all the above mentioned results is that they deal with {\em linear} quantum fields: the time evolution of observables is implemented by a group of linear (symplectic or unitary)  transformations on the phase space, and all the states are quasi-free. 

This means that the problem can be reduced to a question about linear partial differential equations, with boundary conditions on the star boundary.
The Hawking effect emerges from the fact that the star boundary becomes asymptotically characteristic for large times. This leads to an exponentially fast concentration of Klein-Gordon or Dirac wave packets reflected by the star,  which ultimately implies the Hawking effect.

In this paper we investigate the Hawking effect for a toy model of {\em interacting Dirac fermions} in $1+1$ space-time dimensions.
A mathematical discussion of interacting quantum fields is of course  difficult,  because there are few rigorous constructions of interacting quantum fields, even on Minkowski space. 

For Klein-Gordon fields,  there are the well-known constructions of the $P(\varphi)_{2}$ and 
$\varphi^{4}_{3}$ models due to Glimm and Jaffe, which were the main successes of the constructive program from the seventies.
We are not aware of any similar construction on a space-time which describes the exterior of a collapsing star, even when the interaction contains an ultraviolet and space cutoff.

For Dirac fields, the situation looks better, since fermionic fields are bounded, which in some situations allows to construct the interacting dynamics in a purely algebraic setting, independently of the choice of a representation. This is particularly convenient in the situation that we consider, since, even for free Dirac fields, two  Fock representations in the exterior of the star at different times are inequivalent.

\subsection{ A toy model}\label{sec1.2}
To concentrate on the possibly new features introduced by the non-linear interactions and to keep the situation simple and manageable, we restrict ourselves to a toy model of Dirac fermions in $1+1$ space-time dimensions: 

we consider only $2$ components spinors, and the effect of the metric is modeled by  a vector potential. Note that if we forget about the non-linear interaction, our model is essentially identical to the one considered by Bachelot in \cite{Ba1}, after introduction of polar coordinates and suitable spin spherical harmonics.

Let us now briefly describe the model: the space-time is the region:
\[
\cM= \{(t, x)\in \rr^{2}:  x>z(t)\},
\] where $x=z(t)$ is the star boundary. We assume that $z(t)\equiv z(0)$ for $t\leq 0$, i.e. the star is stationary in the past, the collapse starting at $t=0$.  As in \cite{Ba1}  we assume that   $z(t)\sim  -t - A\e^{-2\kappa t}$ for $t\to +\infty$, i.e. the star boundary becomes asymptotically characteristic for large positive times.

The Dirac fields are two-components spinors $\psi(t,x)\in\cc^{2}$, solving (in absence of interaction) the  Dirac equation:
\beq\label{diracdirac}
\left\{
\begin{array}{rl}
&\p_{t}\psi(t,x)+ L\p_{x}\psi(t,x)+ \i V(x) \psi(t,x)=0, \hbox{ in }z>z(t),\\[2mm]
&\psi_{1}(t, z(t))= \lambda(t)\psi_{2}(t, z(t)),
\end{array}\right.
\eeq
where $L=\begin{pmatrix}1&0 \\0&-1\end{pmatrix}$ and $V(x)=V(x)^{*}\in M_{2}(\cc)$ is a matrix-valued potential representing the influence of the metric, with  
\[
V(x)\to 0\hbox{ at }-\infty,\ V(x)\to m\Gamma\hbox{ at }+\infty,
\] $m>0$ is the mass of the field,  and $\Gamma\in M_{2}(\cc)$  satisfies 
\[
\Gamma= \Gamma^{*}, \ \Gamma^{2}= \one, \ \Gamma L + L \Gamma=0.
\]
 The reflection coefficient $\lambda(t)$ equals 
$\big(\frac{1+\dot{z}(t)}{1-\dot{z}(t)}\big)^{1/2}$, so that the  $L^{2}$ norm
\[
 \int_{z(t)}^{+\infty}\| \psi(t,x)\|^{2}_{\cc^{2}}dx
 \] 
is conserved. This implies that if $\ch_{t}:= L^{2}(]z(t), +\infty[; \cc^{2})$, the evolution group $u^{V}(s,t): \ch_{t}\to \ch_{s}$ (see Subsect. \ref{sec2.2}) associated to (\ref{diracdirac}) is unitary, and hence generates a fermionic dynamics 
\[
\tau^{V}(s,t): \CAR(\ch_{t})\to \CAR(\ch_{s}),
\] where $\CAR(\ch)$ is the $\CAR$ $C^{*}-$algebra associated to a Hilbert space $\ch$.

The self-interaction of the Dirac field is described by a perturbation  of the form
\[
I=(\psi^{*}(g)M \psi(g))^{n},
\]
where  $n\geq 2$, $M\in M_{2}(\cc)$ is a selfadjoint matrix, $g\in L^{2}_{\rm comp}(\rr)$ is a compactly supported function.
The associated interacting Dirac fields $\psi^{\rm int }(t,x)$ formally solve the following non-linear Dirac equation:
\beq\label{diracnl}\left\{
\begin{array}{rl}
&\p_{t}\psi^{\rm int }(t,x)+  L\p_{x}\psi^{\rm int }(t,x) +\i V(x)\psi^{\rm int }(t,x)\\[2mm]
-&\i n(\overline{\psi^{\rm int }(t, g)}| M \psi^{\rm int }(t,g))_{\cc^{2}}^{n-1}M \psi^{\rm int }(t,g)g(x)=0,\hbox{ in }x>z(t)\\[2mm]
&\psi_{1}(t, z(t))= \lambda(t)\psi_{2}(t, z(t)),
\end{array}\right.
\eeq
where $\psi^{\rm int }(t,g):= \int \psi^{\rm int }(t,x)\overline{g}(x)dx\in \cc^{2}$. The properties of the interaction which are essential for our analysis are the following:
\ben
\item $I$ is bounded, which allows for a purely algebraic construction of the interacting dynamics $\tauvint(s,t)$;
\item $I$ is even, which is the standard assumption needed to ensure locality,
\item $I$ is localized in a (space) compact region.
\een

\subsection{Results}\label{sec1.3}
Let us now describe the results of the paper. 

The first step is to construct interacting Dirac fields, i.e. to quantize the non-linear Dirac equation (\ref{diracnl}). 

Since we deal with fermions, the interaction term $I$ above is bounded, and one can work in a purely algebraic setting: one can introduce $C^{*}-$algebras $\fA_{t}= \CAR(\ch_{t})$ of observables at time $t$, and it is easy to construct the {\em interacting dynamics} $\tauvint(s,t)$ (see Sect. \ref{sec4}), which is a two parameter group of $*-$isomorphisms from $\fA_{t}$ to $\fA_{s}$ describing the time evolution.

We investigate the Hawking effect in three different situations.
\subsubsection{Hawking effect I} In the first situation  we take an observable at time $t$, localized near the star boundary $x=z(t)$,  i.e. of the form $\alpha^{t}(A)$ for some $A\in \fA_{0}$, where $\alpha^{t}$ is the group of {\em left space translations}. 
In terms of interacting space-time fields $\psi^{\rm int}$, a typical observable would be $\psi^{\rm int}(t , x-t)$,  i.e. a field falling into the black hole along null characteristics. This is the analog  for interacting fields of the situation in  \cite{Ba1}.

To evaluate the  time-evolved state at time $t$ acting on $\alpha^{t}(A)$ we have to evolve $\alpha^{t}(A)$ back to time $0$, which yields
\[
\omega^{V}_{0, {\rm vac}}(\tauvint(0,t)\circ \alpha^{t}(A)),
\]
where  $\omega^{V}_{0, {\rm vac}}$ is the vacuum state at time $t=0$, $\tauvint(t,0)$ is the interacting dynamics. Our goal is to compute the limit of the above quantity when $t\to +\infty$.
We prove  in Thm. \ref{5.0} that  the limit
\beq\label{igor}
\lim_{t\to +\infty}\omega^{V}_{0, {\rm vac}}(\tauvint(0,t)\circ \alpha^{t}(A))= \omega_{\rm H, I}(A)\hbox{ exists},
\eeq
for any $A$ in the  $C^{*}-$algebra $\fA_{0}$.   Let us describe the limiting state $ \omega_{\rm H, I}$, which is close to the one obtained by Bachelot in \cite{Ba1}:

the  algebra $\fA_{0}$ splits into the  ($\zz_{2}-$graded) tensor product $\fA^{\rm l}_{0}\widehat{\otimes}\fA^{\rm r}_{0}$ (see Subsect. \ref{secapp1.2}) of the {\em left/right moving observables}. 

The limit state $\omega_{\rm H, I}$ acts on right moving observables as  a vacuum state (composed with an appropriate wave morphism), while on left moving observables it acts as the thermal state $\omega^{0}_{\infty, \beta}$ at inverse Hawking temperature $\beta= 2\pi \kappa^{-1}$, for the eternal black hole without interaction. 

We also prove a similar result if the initial state $\omega^{V}_{0, {\rm vac}}$ is replaced by another state $\tilde{\omega}$ which is {\em even} and belongs to the {\em folium} of $\omega^{V}_{0, {\rm vac}}$ (see Corollary \ref{5.2}). As example of such a state, one can choose an {\em interacting vacuum state}, whose existence is shown in Subsect. \ref{sec5.2}.

The first situation is graphically summarized in Figure \ref{fig1} below: the grey region is the support of the non-linear self-interaction. The curve $x= z(t)$ is the star boundary. The dashed lines are the (backwards) characteristics for the Dirac equation, starting from the support of an observable at time $T$: left moving characteristics are reflected on the star boundary and asymptotically concentrated when $T\to +\infty$. 
 \begin{figure}[H]
\includegraphics*[scale=1]{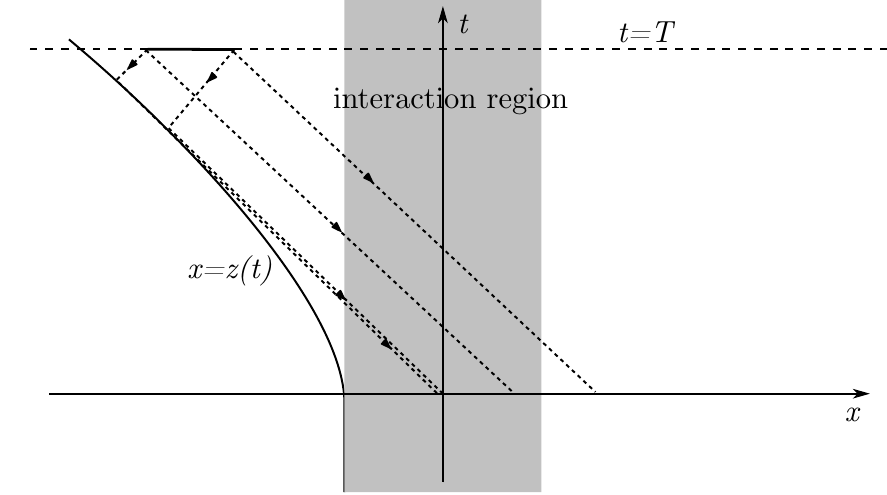}
\caption{Hawking effect I}
\label{fig1}
\end{figure}
\subsubsection{Hawking effect II} In the second situation   the observable  $A$  at time $t$ is  localized near the origin. In terms of  space-time fields, a typical example would be  simply $\psi^{\rm int}(t,x)$.
 This is the analog for interacting fields of the situation  considered by Melnik in  \cite{Me}.

The situation is now more complicated: one has to be sure that the observable $A$, under backwards propagation, will  split into left and right moving parts.   One way to formulate this property is to introduce the
 (future) {\em wave morphism} $\gamma^{\rm  int}_{\infty}$ between the dynamics on the eternal black hole $\tau_{\infty}^{V, {\rm int}}$ and $\tau_{\infty}^{V}$ (see Thm. \ref{6.4}). Then we have to require that  $A$ belongs to  $\gamma^{\rm int}_{\infty}\fA_{\infty}$. Observables outside this $*-$subalgebra will not  see the Hawking effect.

It is easier to formulate our result if we assume  the {\em asymptotic completeness} of $\gamma^{\rm int}_{\infty}$, i.e. that $\gamma^{\rm int}_{\infty}\fA_{\infty}= \fA_{\infty}$: then we prove  in  Thm. \ref{6.13}  that the limit
\beq\label{ogir}
\lim_{t\to +\infty}\omega^{V}_{0, {\rm vac}}(\tauvint(0,t)(A))= \omega_{\rm H, II}(A)\hbox{ exists},
\eeq
for $A$  a {\em local element}  of $\fA_{\infty}$ (i.e. $A\in \fA_{J}$ for some  interval $J\Subset \rr$).

Without assuming asymptotic completeness, we have to restrict ourselves to observables $A\in \gamma^{\rm int}_{\infty}\fA_{\infty}$. Such observables do not necessarily belong to $\fA_{t}$ for $t$ large, i.e. the expression $\tauvint(s,t)(A)$ may have no meaning. Therefore we replace $A$ by $E_{t}A\in \fA_{t}$, where $E_{t}$ is the natural projection $\fA_{\infty}\to \fA_{t}$ (see Remark \ref{igorito}).

Let us now describe the limiting state $\omega_{\rm H, II}$.
Again the 
algebra $\fA_{\infty}$ splits into a tensor product $\CAR(P^{\rm l}\ch_{\infty})\widehat{\otimes}\CAR(P^{\rm r}\ch_{\infty})$ of left/right moving observables (see Subsect. \ref{sec6.1}). In this case elements of $\CAR(P^{\rm l/r}\ch_{\infty})$ are left/right moving only asymptotically for large times. 

On right moving observables the limit state $\omega_{\rm H, II}$ acts again as a vacuum state, composed with a wave morphism. On left moving observables it acts as the thermal state $\omega^{V}_{\infty, \beta}$. In  contrast to case I, the potential term  $V$ is present in the thermal state. 

A similar result holds if we replace the initial state by another even,  state $\tilde{\omega}$ belonging to the folium of $\omega^{V}_{0,{\rm vac}}$, see Corollary \ref{6.14}. However we have now to assume that $\tilde{\omega}$ is invariant under the interacting {\em stationary } dynamics, $\tauvint_{0}$, describing the  interacting Dirac field in the past. 

\begin{figure}[H]
\includegraphics*[scale=1]{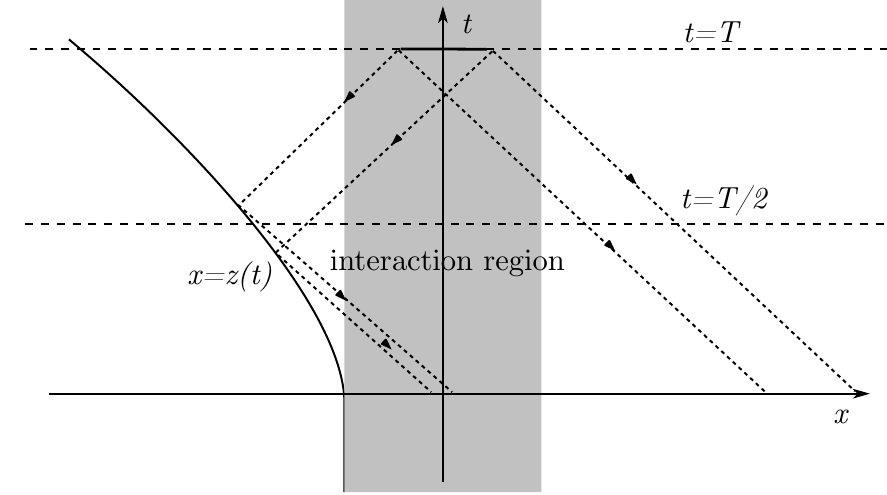}
\caption{ Hawking effect II}
\label{fig2}
\end{figure}

Fig. \ref{fig2} summarizes the second situation, with the same conventions as in Fig. \ref{fig1}: note that left moving characteristics starting at time $T$ from close to the origin, reach the star boundary at time close to $T/2$:  after time $T/2$ the  situation for left moving observables is similar  to case I.
\subsubsection{Hawking effect III}
In the two previous situations, the interaction region is far away from the star boundary: the effect of the self-interaction is decoupled from the effect of the asymptotically caracteristic boundary, which is essential in the Hawking effect.  
 
For an initial observable starting at time $T$ close to the star boundary $z= z(T)$, the Hawking effect (in the free situation),  is essentially due to what happens between the times $T$ and $T-1$, i.e. to the reflection on the asymptotically characteristic star boundary.   Therefore we consider a third situation where the interaction is localized near the star boundary for times $t\in [T-1, T]$.  We consider the following time-dependent interaction 
\[
I_{T}(t)= \one_{[T-1, T]}(t)\alpha^{t}(I),
\]
which is at time $t$ localized near the star boundary $x= z(t)$, and vanishes for $t\not\in [T-1, T].
$
 We denote by $\tilde{\tau}^{V,{\rm int}}_{T}(s,t)$ the dynamics obtained as before by adding to the free dynamics $\tau^{V}(s,t)$ the time-dependent interaction $I_{T}(t)$. We obtain a dynamics depending on the parameter $T$,
which differs from the free dynamics $\tau^{V}(s,t)$ only for $T-1\leq s\leq t\leq T$. 
We show in  Thm. \ref{h5} that the limit
\[
\lim_{T\to \infty}\omega^{V}_{0, {\rm vac}}(\tilde{\tau}^{V, {\rm int}}_{T}(0, T)\circ \alpha^{t}(A))=\omega_{\rm H, III}(A)\hbox{ exists}
\]
for $A\in \fA_{0}$. The limiting state $\omega_{\rm H, III}$ is actually quite explicit, being the pullback of the (free) limiting state $\omega^{\rm free}_{\rm H}$ obtained by Bachelot in \cite{Ba1} by a simple effective interacting dynamics $\hat{\tau}^{0, {\rm int}}_{\infty}(0, 1)$. The dynamics  $\hat{\tau}^{0, {\rm int}}_{\infty}(s, t)$ 
describes  the combined effect of interaction and reflection on the star boundary  between times $T+t$ and $T+s$, in the limit $T\to +\infty$.
The situation is summarized in Fig. \ref{fig3} below.
\begin{figure}[H]
\includegraphics*[scale=1]{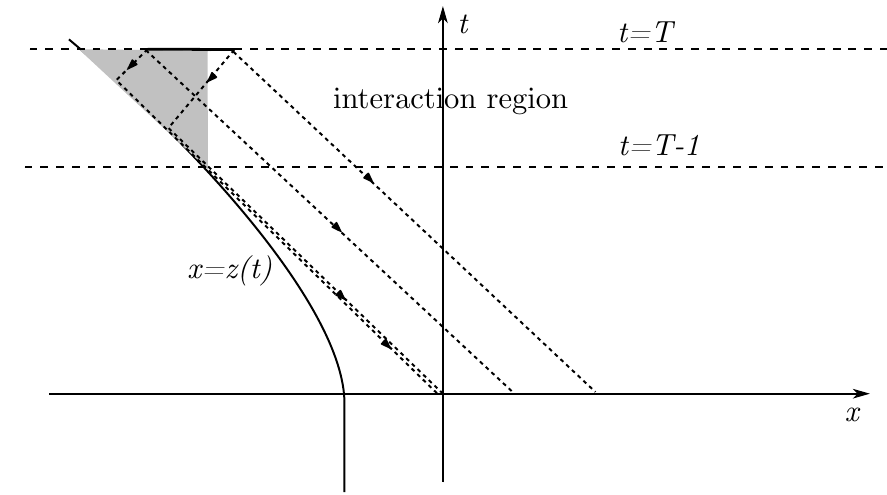}
\caption{ Hawking effect III}
\label{fig3}
\end{figure}
\subsection{Plan of the paper}
Let us now briefly describe the plan of our paper. In Sect. \ref{sec2} we describe our geometrical setup and recall some results of \cite{Ba1} about the linear case.  The corresponding results for quantum dynamics are recalled in Sect. \ref{sec3}. 

In Sect. \ref{sec4} we construct the interacting dynamics in the algebraic, i.e. representation independent setting, by adapting standard perturbation arguments.

 Sect. \ref{sec5} resp. Sect. \ref{sec6}, Sect. \ref{sec7} are devoted to the proof of the Hawking effect in the first, resp. second  and third setup.  
 In Appendix \ref{secapp1} we recall some standard facts about $\CAR$ algebras, the fermionic exponential law and perturbations of $C^{*}-$dynamics.
 
 \subsection{Notations}

If $\ch_{i}$ are Hilbert spaces $i=1,2$  we write $T: \ch_{1}\ \tilde{\to}\ \ch_{2}$   if $T\in B(\ch_{1}, \ch_{2})$ is bijective with bounded inverse.  We will use the same notation if $\fA_{i}$ are $C^{*}-$algebras and $T: \fA_{1}\to \fA_{2}$ is a $*-$isomorphism.

Various objects in the text, like Hilbert spaces, selfadjoint operators, $C^{*}-$algebras, $*-$morphisms or states,  are decorated with sub- and supercripts. As a rule subscripts are used to label a time or a time interval, while superscripts are used to label the various interaction terms,  like $0$ for no interaction, $V$ for interaction potential, or ${\rm int}$ for the non-linear interaction. 
Superscripts ${\rm l/r}$ are also used to denote left/right moving observables. Subscripts ${\rm vac}$ and $\beta$ in states are used to denote vacuum or thermal states, at temperature $\beta^{-1}$.
\section{Classical free dynamics}\label{sec2}\init
In this section we describe our setup and recall some results of \cite{Ba1} about the free classical dynamics. We also collect some additional results which will be important in later sections.

\subsection{Notations and hypotheses}\label{sec2.1}
\subsubsection{Collapsing star}
 We  first recall  the framework of Bachelot \cite{Ba1}, describing a star collapsing to a black hole, in a $1+1$ dimensional space-time. 

The space-time is 
\[
\cM := \{(t,x)\in \rr^{2}:x> z(t)\}
\]
where the {\em star boundary} is  $x=z(t)$ with:
\begin{align}\label{definitionsurface}
&z\in C^2(\mathbb{R}),\nonumber
\\
& z(t)=z(0),\ t\leq0,\nonumber
\\
 &z(t)=-t-Ae^{-2\kappa t}+\zeta(t),\:t\geq 0,
\\
&-1\leq \dot{z}(t)\leq0,\:t\geq0,\nonumber
\end{align}
for  $A,\kappa>0$ and 
\begin{equation}\label{prop_zeta}
|\zeta(t)| + |\dot{\zeta}(t)|\leq Ce^{-4\kappa t},\:t\in\mathbb{R},\:C>0.
\end{equation}
The reflection coefficient on the star boundary is: 
 $$ \lambda(t)=\big(\frac{1+\dot{z}(t)}{1-\dot{z}(t)}\big)^{1/2}.$$
 Without loss of generality we can assume that $z(0)=0$.  The second condition in (\ref{definitionsurface}) means that the collapse start at $t=0$, the star being stationary in the past.

\subsubsection{Dirac operators}\label{diri}
We now define various one dimensional Dirac operators. We set
\[
\begin{array}{rl}
\ch_t&:=L^2(]z(t),+\infty[,\mathbb{C}^2), \; t\in\mathbb{R},\\[2mm]
 \ch_\infty&:=L^2(\mathbb{R},\mathbb{C}^2),\\[2mm]
 \ch_J&:=L^2(J,\mathbb{C}^2),J\Subset\mathbb{R} \text{ interval. }
\end{array}
\]
% We denote by  $\rho_t,t\in\mathbb{R}$ the trace operator:
%\[
%\begin{array}{rl}
%\rho_{t}: &\ H^{1}(\rr, \cc^{2})\ni u\mapsto  u(t)\in \cc^{2},\\[2mm]
%\rho^{*}_{t}:& \cc^{2}\ni v\mapsto \delta_{t}\otimes v\in H^{-1}(\rr, \cc^{2}),
%\end{array}
%\]
%where $H^{p}(\rr, \cc^{2})$ are the usual Sobolev spaces. 

We set
\[
L:= \begin{pmatrix}
1&0 \\
0&-1
\end{pmatrix}
\]
and  fix a matrix-valued potential (representing the influence of the metric):
\[
\rr \ni x \mapsto V(x)\in M_{2}(\cc),  \hbox{ with }V= V^{*}, \ V\in C^{1}(\rr),
\]
and:
\begin{equation}
\label{e.hyp1}
\begin{array}{rl}
|V(x)- V(\infty)| + \langle x\rangle |V'(x)|\in O(\x^{-1-\epsilon}), \ x\to +\infty,\\[2mm]
| V(x)| + \x |V'(x)|\in O(\x^{-2- \epsilon}), ÷ x\to -\infty,
\end{array}
\end{equation}
for some $\epsilon>0$. We assume that 
\[
V_{\infty}= m \Gamma, \ \Gamma\in M_{2}(\cc), 
\]
where $m>0$ is the mass of the field and
\[
\Gamma= \Gamma^{*}, \Gamma^{2}= \one, \ \Gamma L + L \Gamma=0.
\]

Let us now introduce Dirac operators; We set:
\begin{equation}
   b^V_t:=\i L\partial_x-V(x)  \hbox{ acting on }\ch_{t},
\end{equation}
with domain \[
\Dom b^V_t=\{u\in H^1(]z(t),+\infty[,\mathbb{C}^2)\ : \ u_{1}(z(t))= \lambda(t)u_{2}(z(t))\},
\]
and:\begin{equation}
   b^V_\infty:=\i L\partial_x-V(x)  \hbox{ acting on }\ch_{\infty}
\end{equation}
 with domain 
 \[
 \Dom b^V_\infty=H^1(\mathbb{R},\mathbb{C}^2).
 \]
 \subsection{Classical  free dynamics}\label{sec2.2}
 The classical  free dynamics is generated by the following Dirac equation:
 \begin{equation}
\label{e2.1}
\left\{
\begin{array}{rl}
&\p_{s}\psi(s,x)+  L\p_{x}\psi(s,x) +\i V(x)\psi(s,x)=0,\hbox{ in }x> z(s),\ s\in \rr,\\[2mm]
&\psi_{1}(s, z(s))= \lambda(s)\psi_{2}(s, z(s)), \ s\in \rr,\\[2mm]
&\psi(t, x)= \psi(x), \hbox{ in }x>z(t).
\end{array}
\right.
\end{equation}
In this subsection we recall some results  of \cite{Ba1}, about the existence and properties of solutions of (\ref{e2.1}).

\begin{definition}\label{2.1}
 A  $\{u(s,t)\}_{s,t\in \rr}$ with values in $B(\ch_{t}, \ch_{s})$ is called a (two-parameter) {\em propagator} if:
 \[
\begin{array}{rl}
i)&u(s,t)\in U(\ch_{t}, \ch_{s}), \\[2mm]
ii)&u(t,t)= \one, \ t\in \rr,\\[2mm]
iii)&u(s,t')u(t', t)= u(s,t), \ s,t',t\in \rr,\\[2mm]
iv)& \forall \ (s_{0}, t_{0})\in \rr^{2}, \ \forall J\Subset ]z(t_{0}), +\infty[\ \forall f\in \ch_{J}\hbox{ the map}\\[2mm]
&(s,t)\mapsto u(s,t)f\in \ch_{\infty}\hbox{ is continuous at }(s_{0}, t_{0}).
\end{array}
\]
\end{definition}
In the above definition we denoted by $U(\ch_{t}, \ch_{s})$ the group of unitary operators from $\ch_{t}$ to $\ch_{s}$.

Note that condition {\it iv)} is the appropriate replacement for the strong continuity of $(s,t)\mapsto u(s,t)$ in the case $\ch_{t}\equiv \ch$.

The following result can be found in \cite{Ba1}.
\begin{theoreme}\label{2.2}
Assume the hypotheses in Subsect. \ref{sec2.1}. Then 
 there exists a unique  propagator $u^{V}(s,t)\in B(\ch_{t}, \ch_{s})$ such that:
 \[
\begin{array}{rl}
&u^{V}(s, t): \Dom b^{V}_{t}\to \Dom b^{V}_{s}, \ s, t\in \rr,\\[2mm]
&\p_{s}u^{V}(s,t)= \i b^{V}_{s} u^{V}(s,t)\hbox{ on }\Dom b^{V}_{t},\\[2mm]
&\p_{t}u^{V}(s,t)= -\i u^{V}(s,t)b^{V}_{t}\hbox{ on }\Dom b^{V}_{t}.
\end{array}
\]
\end{theoreme}
It follows that if $\psi\in \Dom b^{V}_{t}$, then $\psi(s,x)= u^{V}(s,t)\psi(x)$ solves (\ref{e2.1}) in the strong sense.
 For the Dirac equation without boundary condition we will set accordingly:
 \[
u^{V}_{\infty}(s,t):= \e^{\i (s-t)b^{V}_{\infty}}\in U(\ch_{\infty}, \ch_{\infty}).
\]
\subsection{Additional results}\label{sec2.3}
In this subsection we collect some known results from Bachelot \cite{Ba1} about the classical dynamics $u^{V}(s,t)$.
For {\em free} Dirac fields outside of a collapsing star, they are sufficient to obtain a proof of the Hawking effect, as done in \cite{Ba1}.
In the toy model of interacting Dirac fields that we consider, they will also be important.

We first define the {\em left translations}:
\begin{definition}
 \label{2.5} If $f\in \ch_{\infty}$, we set $f^{t}(\cdot):= f(\cdot +t)\in \ch_{\infty}$.
\end{definition}
\subsubsection{Finite propagation speed}
We first  collect some properties of finite propagation speed  for $u^{V}(s,t)$ and $u^{V}_{\infty}(s,t)$.
\begin{proposition}
 \label{2.3}
\ben
\item  if $\supp f\subset [R, +\infty[$ then $\supp u^{V}(s,t)f\subset [R+ |t-s|,+\infty[$;
\item if $\supp f\subset [a,b]$ then $\supp u^{V}_{\infty}(s,t)f\subset [a- |t-s|, b+ |t-s|]$;
\item if $\supp f\subset [0, R]$ then $\supp u^{V}(s,t)f^{t}\subset [z(s), R-s]$ for all $s\leq t$.

\een 
\end{proposition}
\proof  the proof of (1) can be found in \cite{Ba1}. (2) follows from classical arguments, see e.g. \cite{CP82}. (3) is shown in \cite[Proof of Thm. VI.5]{Ba1}. \qed

Statement (2) of Prop. \ref{2.3} and the uniqueness in Thm. \ref{2.2} imply the following fact:
\begin{proposition}
 \label{2.4}
 Let $J\Subset\rr$ an interval. Then there exists $c\geq 0$ such that
 \[
u^{V}(s,t)f= u^{V}_{\infty}(s,t)f, \ \forall \ f\in \ch_{J}, \ c+ t/2\leq s\leq t.
\]
\end{proposition}
\subsubsection{Scattering results}
One can split $\ch_{t}$ as direct sum:
\[
\ch_{t}= \ch_{t}^{\rm l}\oplus \ch^{\rm r}_{t},
\]
for
\begin{equation}
\label{e2.2}
\ch^{\rm l}_{t}:= \{f=(f_{1}, f_{2})\in \ch_{t}\ : \ f_{2}=0\},\
 \ch^{\rm r}_{t}:= \{f=(f_{1}, f_{2})\in \ch_{t}\ : \ f_{1}=0\}.
\end{equation}
If $f\in \ch_{t}$, we denote by $f^{\rm l/r}$ its orthogonal projection on $\ch_{t}^{\rm l/r}$.

If $V\equiv 0$ we easily see that:
\begin{equation}
\label{e2.3}
u^{0}_{\infty}(0,t)f=f^{t}, \ f\in \ch_{\infty}^{\rm l},\
u^{0}_{\infty}(t,0)f= f^{t},\   f\in \ch_{\infty}^{\rm r}.
\end{equation}
\begin{proposition}\label{2.6}
 The strong limit
 \[
w^{\rm r}:= \slim_{t\to+ \infty} u^{V}(0,t)u^{0}_{\infty}(t,0)
\]
exists on $\ch^{d}_{\infty}$.
\end{proposition}
 \proof See \cite[Prop. VI.4]{Ba1}. \qed
 
\begin{proposition}
 \label{2.6bis}
 \[
\wlim_{t\to +\infty} u^{V}(0, t)f^{t}=0, \ \forall \ f\in \ch^{\rm l}_{0}.
\]
\end{proposition}
\proof We follow some arguments in \cite{Ba1}. By density we can assume that  $f\in \ch^{\rm l}_{0}$  is compactly supported. We write for  $0\leq T\leq t$:
\[
\begin{array}{rl}
&\|u^{V}(T,t)f^{t}- u^{0}(T, t)f^{t}\|=\|u^{0}(t, T)u^{V}(T,t)f^{t}- f^{t}\|\\[2mm]
=&\|\int_{T}^{t} u^{0}(t, \sigma) V u^{V}(\sigma, t)f^{t} d\sigma\|\leq \int_{T}^{t}\| V u^{V}(\sigma, t)f^{t}\| d\sigma.
\end{array}
\]
By Prop. \ref{2.3} (3) we know that $\supp u^{V}(\sigma, t)f^{t}\subset [z(\sigma), R- \sigma]$ for some $R\geq 0$, hence by hypothesis (\ref{e.hyp1}) we have $\| V u^{V}(\sigma,t)f^{t}\|\in O(\langle \sigma\rangle ^{-2- \epsilon})$. It follows that 
\beq\label{e2.00}
\lim_{T\to+\infty}\sup_{T\leq t}\| u^{V}(T,t)f^{t}- u^{0}(T, t)f^{t}\|=0.
\eeq
Next we write
\[
\begin{array}{rl}
u^{V}(0,t)f^{t}=& u^{V}(0, T)u^{0}(T,t)f^{t}+ u^{V}(0, T)\left(u^{V}(T, t)f^{t}- U^{0}(T, t)f^{t}\right)\\[2mm]
=& u^{V}(0, T)u^{0}(T,0)u^{0}(0, t)f^{t}+ u^{V}(0, T)\left(u^{V}(T, t)f^{t}- U^{0}(T, t)f^{t}\right).
\end{array}
\]
We know from \cite[Lemma VI.8]{Ba1} that $\wlim_{t\to +\infty}u^{0}(0, t)f^{t}=0$.  Using (\ref{e2.00}) and an $\epsilon/2$ argument we obtain the proposition. \qed

 \subsubsection{Limits of quasi-free states}
 The following theorem is the key result of \cite{Ba1}.
\begin{theoreme}\label{2.7}
 For $f\in \ch^{\rm l}_{0}$ one has:
 \[
\lim_{t\to +\infty}(u^{V}(0,t)f^{t}| \one_{\rr^{+}}(b_{0}^{V})u^{V}(0,t)f^{t}) = (f| (\one+ \e^{- 2\pi\kappa^{-1} b^{0}_{\infty}})^{-1}f).
\]
\end{theoreme}

The analogous result for $f\in \ch_{0}^{\rm r}$ follows immediately from Prop. \ref{2.6} and (\ref{e2.3}).
\begin{proposition}\label{2.8}
 For $f\in \ch^{\rm r}_{0}$ one has:
 \[
\lim_{t\to +\infty}(u^{V}(0,t)f^{t}| \one_{\rr^{+}}(b_{0}^{V})u^{V}(0,t)f^{t}) =(w^{\rm r}f| \one_{\rr^{+}}(b_{0}^{V})w^{\rm r}f).
\]
\end{proposition}
We recall that $(f|\one_{\rr^{+}}(b_{0}^{V})f)$ is the covariance of the quasi-free vacuum state for the Dirac field in the exterior of the star at $t=0$, while $(f| (\one+ \e^{- 2\pi\kappa^{-1} b^{0}_{\infty}})^{-1}f)$ is the covariance of the thermal state at Hawking temperature $\kappa/2\pi$ near the black hole horizon.

\section{Free quantum  dynamics}\label{sec3}\init

In this section we define the free quantum dynamics corresponding to the classical dynamics constructed in Subsect. \ref{sec2.2}.

Let us first introduce some notation. For $t\geq 0$ we set ${\mathfrak A}_{t}:= \CAR(\ch_{t})$, $ \fA_{\infty}:= \CAR(\ch_{\infty})$ and for an interval  $J\Subset \rr$, $\fA_{J}:= \CAR(\ch_{J})$ (see Subsect. \ref{secapp1.1}).
Note that $\fA_{t}, \ \fA_{J}\subset \fA_{\infty}$ isometrically. 

We start by a definition  analogous to Def. \ref{2.1}.
\begin{definition}
 \label{3.1} A family $\{\tau(s,t)\}_{s,t\in \rr}$ is a (two-parameter) {\em quantum dynamics} if:
 \[
\begin{array}{rl}
i)& \tau(s,t):\ \fA_{t}\ \tilde{\to}\  \fA_{s},\\[2mm]
ii)& \tau(t,t)= \one, t\in \rr,\\[2mm]
iii)& \tau(s,t')\tau(t',t)= \tau(s,t), \ s, t',t\in \rr,\\[2mm]
iv)& \forall \ (s_{0}, t_{0})\in \rr^{2}, \ \forall J\Subset ]z(t_{0}), +\infty[\ \forall A\in \fA_{J}\hbox{ the map}\\[2mm]
&(s,t)\mapsto \tau(s,t)A\in \fA_{\infty}\hbox{ is continuous at }(s_{0}, t_{0}).
\end{array}
\]
\end{definition}

Since $u^{V}(t,s)$ is a propagator, it generates a (free) quantum dynamics $\tau^{V}(s,t)$.
\begin{definition}
 \label{3.2}
 We denote by $\tau^{V}(s,t)$ the quantum dynamics defined by:
 \[
\tau^{V}(s,t)(\psi^{(*)}(f)):= \psi^{(*)}(u^{V}(s,t)f), \ f\in \ch_{t}.
\]
\end{definition}
Similarly we define the quantum dynamics $\tau^{0}(s,t)$, $\tau^{V}_{\infty}(s,t)$ associated to $u^{0}(s,t)$ and $u^{V}_{\infty}(s,t)$. 

Note that $\tau^{V}_{\infty}(s,t)$ is a  {\em stationary }quantum dynamics on $\fA_{\infty}$, i.e. $\tau^{V}_{\infty}(s,t)= \tau^{V}_{\infty}(s+t', t+t')$, for all $s,t',t\in \rr$.

We also define the (one-parameter)  dynamics $\alpha^{t}$ on $\fA_{\infty}$ defined by
\beq\label{def-de-alpha}
 \alpha^{t}(\psi^{(*)}(f)):= \psi^{(*)}(f^{t}), \ f\in \ch_{\infty}.
\eeq
The properties of propagators recalled in Subsect. \ref{sec2.3} immediately carry over to  quantum dynamics. For example the following fact follows from Prop. \ref{2.4}.
\begin{lemma}
 \label{3.3}
 Let $J\Subset \rr$ an interval.  Then there exists $c\geq 0$ such that
 \[
\tau^{V}(s,t)A= \tau^{V}_{\infty}(s,t)A, \ \forall \ A\in \fA_{J}, \ c+t/2\leq s\leq t.
\]
\end{lemma}
\section{Interacting quantum dynamics}\label{sec4}\init

In this section we construct the interacting dynamics $\tauvint(s,t)$ that we will consider in the sequel. It will be obtained by perturbing  the free dynamics $\tau^{V}(s,t)$ by a bounded interaction term $I$ localized in a bounded region of space. As usual, since we consider fermionic fields , interacting dynamics can be constructed at the algebraic level.

Formally the construction of the interacting dynamics $\tauvint(s,t)$ defined in Def. \ref{4.2} corresponds to the quantization of the following non-linear Dirac equation:
\beq\label{nld}
\left\{
\begin{array}{rl}
&\p_{s}\psi(s,x)+  L\p_{x}\psi(s,x) +\i V(x)\psi(s,x)\\[2mm]
-&\i n(\overline{\psi(s, g)}| M \psi(s,g))_{\cc^{2}}^{n-1}M \psi(s,g)g(x)=0,\\[2mm]
&\psi_{1}(s, z(s))= \lambda(s)\psi_{2}(s, z(s)), \ s\in \rr,\\[2mm]
&\psi(t, x)= \psi(x), \hbox{ in }x>z(t),
\end{array}
\right.
\eeq
where $\psi(s,g):= \int \psi(s,x)\overline{g}(x)dx\in \cc^{2}$, $M\in M_{2}(\cc)$ is a selfadjoint matrix and $g\in L^{2}(J)$ for some $J\Subset \rr$ is a compactly supported function.
\subsection{Construction of the interacting dynamics}\label{sec4.1}
\begin{definition}\label{4.1}
 Let $M\in M_{2}(\cc)$ with $M = M^{*}$. We set for $g\in L^{2}(\rr)$:
 \[
\psi^{*}(g)M \psi(g):= \sum_{i=1}^{2}\psi^{*}(g\otimes e_{i})\lambda_{i}\psi(g\otimes e_{i})\in \fA_{\infty},
\]
where $M= \sum_{i=1}^{2} \lambda_{i}|e_{i}) (e_{i}|$ (i.e. $(e_{1}, e_{2})$ is a basis of eigenvectors of $M$).

\end{definition}

We fix $g\in L^{2}(J)$ for an interval $J\Subset ]z(0), +\infty[$ , $2\leq n\in \nn$ and set:
\begin{equation}
\label{e4.1}
I:= (\psi^{*}(g)M \psi(g))^{n}\in \CAR_{0}(\ch_{0}).
\end{equation}
The interaction term $I$ represent a localized, even, self-interaction of the Dirac field in $\cM$.
\begin{remark}
 All the results below extend immediately to the case when $I$ is replaced by a finite sum  of $I_{k}$, associated to matrices $M_{k}$ and compactly supported space-cutoffs $g_{k}$. The only important properties of $I$ is that it should be even and localized.
\end{remark}

For later use we state the following fact, which follows immediately from the CAR and the fact that $I\in \CAR_{0}(\ch_{\infty})$.
\begin{lemma}
 \label{4.0}
 Let $B=\prod_{i=1}^{n}\psi^{(*)}(f_{i})$. Then there exists $C_{n}$ such that
 \begin{equation}
\label{e4.0}
\| [I, B]\| \leq C_{n}\prod_{i=1}^{n}\| f_{i}\| \sum_{i=1}^{n}| (g| f_{i})|.
\end{equation}
\end{lemma}

Using the results of Subsect. \ref{secapp.4} we can now construct the {\em interacting dynamics} $\{\tauvint(s,t)\}_{s,t\in \rr}$.
\begin{definition}\label{4.2}
 Let $I$ be as in (\ref{e4.1}) , $I(s,t):= \tau^{V}(s,t)I$ and $R(s,t):=R_{s}(s,t)\in U(\fA_{s})$  be obtained  as in Prop.  \ref{app2.2}.
 We set:
 \beq\label{e4.1bis}
\tauvint(s,t)(A):= R(s,t)\tau^{V}(s,t)(A) R(s,t)^{*}, \ A\in \fA_{t}, \ s,t\in \rr
\eeq
 Then by Prop. \ref{app2.2} $\{\tauvint(s,t)\}_{s,t\in \rr}$ is a dynamics, called the {\em interacting quantum dynamics}. 
 \end{definition}
 For the convenience of the reader, we recall that $R(s,t)\in \fA_{s}$ solves:
 \beq\label{e4.00}
\left\{
\begin{array}{rl}
\p_{\sigma}R(s, \sigma)=& -\i R(s, \sigma) I(s,\sigma),\\[2mm]
R(s,s)=&\one.
\end{array}
\right.
\eeq

We can also define the corresponding interacting dynamics without boundary conditions, acting on $\fA_{\infty}$.
We set $I_{\infty}(s,t):= \tau^{V}_{\infty}(s,t)(I)$ and define  $R_{\infty}(s,t)\in U(\fA_{\infty})$ as above and:
\beq\label{e4.2}
\tauvint_{\infty}(s,t)(A):= R_{\infty}(s,t)\tau^{V}_{\infty}(s,t) (A)R_{\infty}(s,t)^{*}, \ A\in \fA_{\infty}, \s,t\in \rr.
\eeq
Again $\tauvint_{\infty}(s,t)$ is stationary.
\begin{remark}
 Let us faithfully represent $\fA_{\infty}= \CAR(\ch_{\infty})$ in the fermionic Fock space $\Gamma_{\rm a}(\ch_{\infty})$ (see Subsect. \ref{app1.1}) by the Fock representation $\pi_{F}$. Then $\tau^{V}_{\infty}(s,t)$ is implemented in the Fock representation by the unitary group $\e^{\i (s-t)H^{V}_{\infty}}$, where $H_{V}^{\infty}= \d\Gamma(b^{V}_{\infty})$ is the second quantization of $b^{V}_{\infty}$. The dynamics $\tauvint_{\infty}(s,t)$ is implemented by $\e^{\i (s-t)H^{V,{\rm int}}_{\infty}}$ for $H^{V,{\rm int}}_{\infty}= H^{V}_{\infty}+ \pi_{F}(I)$.
\end{remark}
\subsection{Properties of $\tauvint(s,t)$}\label{sec4.2}

\begin{lemma}
 \label{4.3}
There exists $c\geq 0$ such that:
 \[
 R_{\infty}(s,t)= R(s,t), \ c+t/2\leq s\leq t.
\]
\end{lemma}
\proof  The interaction $I$ defined in (\ref{e4.1}) belongs to $\fA_{J}$ for some interval $J\Subset \rr$. We apply then Lemma \ref{3.3} to each term in the series defining $R(s,t)$, see Lemma \ref{app2.1}. \qed

\begin{lemma}
 \label{4.4}
 Let $J\Subset\rr$ an interval. Then there exists $c\geq 0$ such that
 \[
\tauvint(s,t)(A)= \tauvint_{\infty}(s,t)(A), \ \forall \ A\in \fA_{J}, \ c+t/2\leq s\leq t.
\]
\end{lemma}
\proof It suffices to apply Lemmas \ref{4.3} and \ref{3.3} to the definition of $\tauvint$, $\tauvint_{\infty}$. \qed

\section{Hawking effect I}\label{sec5}\init
In this section we study the Hawking effect in the situation referred to as case I in the introduction (see Subsect. \ref{sec1.3}). For $t\in \rr$ we set $\fA_{t}^{\rm l/r}:= \CAR(\ch_{t}^{\rm l/r})\subset \fA_{t}$, called the {\em left/right moving observables}.

The algebra $\fA_{0}$ splits into a twisted tensor product of the left/right moving $\CAR$ algebras $\fA^{\rm l/r}_{0}$. The first step consists in studying the evolution $\tauvint(0,t)\circ \alpha^{t}$ on left/right moving observables.
\subsection{Left propagation}\label{sec4.3}
\begin{proposition}
 \label{4.5}
 Let $A\in \fA^{\rm l}_{0}$. Then 
 \[
\lim_{t\to+\infty}\tauvint(0,t)\circ \alpha^{t}(A)- \tau^{V}(0,t)\circ \alpha^{t}(A)=0.
\]
\end{proposition}
To prove Prop. \ref{4.5}, we will need the following lemma.

\begin{lemma}
 \label{4.6}
 For any $\epsilon>0$ and $A\in \fA_{0}$, there exists $T$ such that
 \[
\sup_{t\geq T}\| \tauvint(T, t)\circ \alpha^{t}(A)- \tauv(T, t)\circ \alpha^{t}(A)\|\leq \epsilon.
\]
\end{lemma}

\proof Let us set $A(s,t)= \tauv(s,t)\circ \alpha^{t}(A)$ to simplify notation. 
We first claim that
\beq\label{e4.5}
\| \tauvint(s,t)\circ \alpha^{t}(A)- A(s,t)\|\leq\int_{s}^{t}\| [I, A(\sigma, t)]\|d\sigma.
\eeq
Let us prove (\ref{e4.5}).
By Def. \ref{4.2} we have:
\[
\begin{array}{rl}
&\tauvint(s,t)\circ \alpha^{t}(A)- A(s,t)= R(s,t) A(s,t) R(s,t)^{*}-A(s,t)\\[2mm]
=& [R(s,t), A(s,t)] R^{*}(s,t),
\end{array}
\]
using that $R(s,t)$ is unitary. Set :
\[
F_{s,t}(\sigma):= [R(s, \sigma), A(s,t)], \ G_{s,t}(\sigma):= [I(s,\sigma), A(s,t)].
\]
We note first that
\[
G_{s,t}(\sigma)= [\tauv(s, \sigma)(I), \tauv(s,t)\circ \alpha^{t}(A)]= \tauv(s, \sigma)([I, A(\sigma, t)]),
\]
using that $\tauv$ is an homomorphism. Since $\tauv$ is isometric, we have
\begin{equation}
\label{e4.6}
\| G_{s,t}(\sigma)\|= \|[I,A(\sigma, t)]\|.
\end{equation}
Recalling that $R(s,\sigma)$ solves
\[
\left\{
\begin{array}{rl}
\p_{\sigma}R(s, \sigma)=& -\i R(s, \sigma) I(s,\sigma),\\[2mm]
R(s,s)=&\one,
\end{array}
\right.
\]
we see next  that $F_{s,t}(\cdot)$ solves the equation:
\beq\label{e4.3}
\left\{\begin{array}{rl}
\p_{\sigma}F_{s,t}(\sigma)=& -\i F_{s,t}(\sigma)I(s,\sigma)- \i R(s, \sigma)G_{s,t}(\sigma),\\[2mm]
F_{s,t}(s)=&0,
\end{array}\right.
\eeq
which clearly has a unique solution. We look for $F_{s,t}(\sigma)$ of the form $F_{s,t}(\sigma)= H_{s,t}(\sigma)R(s,\sigma)$. We obtain the equation:
\begin{equation}
\label{e4.4}
\left\{
\begin{array}{rl}
\p_{\sigma} H_{s,t}(\sigma)=& -\i R(s,\sigma)G_{s,t}(\sigma)R(s, \sigma)^{*},\\[2mm]
H_{s,t}(s)=&0.
\end{array}
\right.
\end{equation}
Since $R(s,\sigma)$ is unitary we obtain
\[\begin{array}{rl}
&\| \tauvint(s,t)\circ \alpha^{t}(A)- A(s,t)\|= \| F_{s,t}(t)\| = \| H_{s,t}(t)\|\\[2mm]
\leq& \int_{s}^{t}\|G_{s,t}(\sigma)\| d \sigma=\int_{s}^{t}\| [I, A(\sigma, t)]\|d\sigma,
\end{array}
\]
which proves (\ref{e4.5}). 

We can now complete the proof of the lemma. 
Assume first that $A$ belongs to $\CAR_{\rm alg}(\ch_{J})$ for some interval $J\subset [0, R]$ (recall that $z(0)=0$). By linearity we may assume that $A=\prod_{i=1}^{n} \psi^{(*)}(f_{i})$ with $\supp f_{i}\subset [0, R]$. By Prop. \ref{2.3}  (3) we know that $\supp u^{V}(\sigma,t)f_{i}^{t}\subset [z(\sigma), -\sigma +R]$ hence for $\sigma\geq \sigma(J)$ we have $[I, A(\sigma, t)] =0$ by Lemma \ref{4.0},
hence 
\beq\label{e4.7}
\tauvint(s,t)(A)- A(s,t)=0, \ \sigma(J)\leq s\leq t, \ A\in \CAR_{\rm alg}(\ch_{J}).
\eeq
Let now $A\in \fA_{0}$ and $\epsilon>0$. By density we can choose $J$ as above and $\tilde{A}\in \CAR_{\rm alg}(\ch_{J})$ such that $\| A- \tilde{A}\|\leq \epsilon/2$. Applying (\ref{e4.7}) to $\tilde{A}$ we obtain $T= \sigma(J)$ such that
\[
\sup_{T\leq t}\| \tauvint(T,t)\circ \alpha^{t}(A)- A(T,t)\|\leq \epsilon.
\]
This completes the proof of the lemma. \qed

{\bf Proof of Prop. \ref{4.5}.} Let $A\in \fA_{0}^{\rm l}$. Again let us set $A(s,t)= \tauv(s,t)\circ \alpha^{t}(A)$, so that we need to show that
\[
\lim_{t\to+\infty}\tauvint(0,t)\circ \alpha^{t}(A)- A(0,t)=0.
\] We fix $\epsilon>0$ and $T$ as in Lemma \ref{4.6}.   We have:
\[
\begin{array}{rl}
\tauvint(0,t)\circ \alpha^{t}(A)=&\tauvint(0, T)\circ \tauvint(T, t)\circ \alpha^{t}(A)\\[2mm]
=& R(0, T)\tauv(0,T)\circ \tauv(T, t)\circ\alpha^{t}(A) R(0,T)^{*}\\[2mm]
=& R(0,T)A(0,t)R(0,T)+ O(\epsilon), 
\end{array}
\]
by Lemma \ref{4.6}.   By (\ref{e4.00}) we have:
\[
\p_{\sigma}( R(0,\sigma)B R(0, \sigma)^{*})= -\i R(0, \sigma)[I(0, \sigma), B]R(0, \sigma)^{*},\ B\in \fA_{0}
\]
hence:
\[
R(0, T)A(t)R(0, T)^{*}-A(0,t)= -\i \int_{0}^{T}R(0, \sigma)[I(0, \sigma), A(0,t)]R(0, \sigma)^{*}d\sigma.
\]
By (\ref{e4.6}) for $s=0$ we have $\| [I(0,\sigma), A(0,t)]\|= \| [I, A(\sigma, t)]\|$. To complete the proof of the proposition, it suffices to show that
\begin{equation}
\label{e4.8}
\lim_{t\to+\infty}[I, A(\sigma, t)]=0, \ \forall \  \sigma\geq 0.
\end{equation}
Since $\|A(\sigma, t)\|= \|A\|$, it suffices by density and linearity to prove (\ref{e4.8}) if $A=\prod_{i=1}^{n}\psi^{(*)}(f_{i})$ for $f_{i}\in \ch^{\rm l}_{0}$ with compact support. 
By Lemma \ref{4.0} it suffices hence to prove that 
\[
\wlim_{t\to +\infty}u^{V}(0,t)f_{i}^{t}=0.
\]
But this follows from Prop. \ref{2.6bis}.  This completes the proof of the proposition. \qed

\subsection{Right propagation}\label{sec4.4}
\begin{proposition}
 \label{4.7}
 The strong limit
 \[
\slim_{t\to +\infty}\tauvint(0,t)\circ \alpha^{t}=: \gamma^{\rm r, int}
\]
exists on $\fA^{\rm r}_{0}$.
\end{proposition}
Before proving the proposition, let us note that $\gamma^{\rm r, int}$ is an even homomorphism  (see Subsect. \ref{app1.1}).
\begin{lemma}
 \label{4.8}
 The homomorphism $\gamma^{\rm r, int}$ is even i.e. $P\circ \gamma^{\rm r, int}= \gamma^{\rm r, int}\circ P$. 
\end{lemma}
\proof $\alpha^{t}$ is even, so it suffices to prove that $\tauvint(0,t)$ is even. This follows if we prove that $R(s,t)\in \CAR_{0}(\ch_{s})$.  We note  that $R(s, \sigma)$ and  $PR(s, \sigma)$ solve the same differential equation, using that $I$ is even. \qed

{\bf Proof of Prop. \ref{4.7}.} Let $A\in \fA^{\rm r}_{0}$.  By (\ref{e2.3}) we have $\alpha^{t}= \tau^{0}_{\infty}(t,0)$ on $\fA^{\rm r}_{0}$. Therefore we will be able to prove the proposition by the Cook argument. We will first prove that
\begin{equation}
\label{e4.9}
\lim_{t\to+\infty} \tauv(0,t)\circ \tau^{0}_{\infty}(t,0)(A)=: \gamma_{0}^{\rm r}(A), \ A\in \fA_{0}^{\rm r}.
\end{equation}
exists, and then that
\begin{equation}
\label{e4.10}
\slim_{t\to +\infty}\tauvint(0,t)\circ \tauv(t,0)(A), \ A\in \gamma^{\rm r}_{0}\fA^{\rm r}_{0}.
\end{equation}
exists. Let us first prove (\ref{e4.9}). Since $\tauv(0,t)$ and $\tau^{0}_{\infty}(t,0)$ are free dynamics, this follows from  Prop. \ref{2.6} which states that:
\beq\label{e4.11}
\lim_{t\to +\infty}u^{V}(0,t)u^{0}_{\infty}(t,0)f= w^{\rm r}_{0}f
\eeq
exists for $f\in\ch^{\rm r}_{0}$. It follows that 
\[
 \gamma^{\rm r}_{0}(\psi^{(*)}(f))= \psi^{(*)}(w_{0}^{\rm r}f), \ f\in \ch_{0}^{\rm r}.
\]
To prove (\ref{e4.10}) we will need some estimates on the speed of convergence in (\ref{e4.11}), for well chosen initial data.

%icirecheck
Assume that $f\in \ch^{\rm r}_{0}$ is smooth with compact support.   Then $u^{0}_{\infty}(t,0)f= f^{t}\equiv 0$ near $x= z(t)$ hence $u^{0}_{\infty}(t,0)f\in \Dom b^{V}_{t}$. It follows that:
\[
\begin{array}{rl}
\p_{t}u^{V}(0,t)u^{0}_{\infty}(t,0)f= \i u^{V}(0,t)(b^{0}_{\infty}- b^{V}_{t})u^{0}_{\infty}(t,0)f= \i  u^{V}(0,t)V f^{t}.
\end{array}
\]
From hypothesis (\ref{e.hyp1}) we obtain that $\| V f^{t}\|\in O(t^{-2- \epsilon})$ hence by integrating from $t$ to $+\infty$, we obtain:
\begin{equation}
\label{e4.12}
w^{\rm r}_{0}f- u^{V}(0,t)u^{0}_{\infty}(t,0)f\in O(t^{-1- \epsilon}).
\end{equation}
Let us now prove (\ref{e4.10}).  By linearity, density and using that $\tauvint(0,t)$ and $\tau^{V}(t,0)$ are isomorphisms, we can assume that 
 $A= \psi^{(*)}(w_{0}^{\rm r}f)$ for $f\in \ch^{\rm r}_{0}$ smooth with compact support. 
We have 
\[
\begin{array}{rl}
&\tauvint(0,t)\circ \tauv(t,0)(A)=R(0,t)\tauv(0,t)\circ \tauv(t,0)(A) R(0,t)^{*}\\[2mm]
=&R(0,t)AR(0,t)^{*}.
\end{array}
\]
We apply once more the Cook argument and compute
\[
\p_{t}R(0,t)\gamma^{\rm r}_{0}(A)R(0,t)^{*}= -\i R(0,t)[I(0,t),A]R(0,t)^{*}.
\]
As before 
\[
\begin{array}{rl}
&\| [I(0,t), A]\|= \|[I, \tauv(t,0)(A)]\|= \|[I, \psi^{(*)}(u^{V}(t,0)w^{\rm r}_{0}f)]\|\\[2mm]
=& \|  [I, \psi^{(*)}(u^{0}(t,0)f] \|+ O(t^{-1- \epsilon}),
\end{array}
\]
by (\ref{e4.12}). Since $f$ has compact support, and $u^{0}(t,0)f=f^{t}$, we obtain that $[I, u^{0}(t,0)f]=0$ for $t$
 large enough. Therefore $\| \p_{t}R(0,t)\gamma^{\rm r}_{0}(A)R(0,t)^{*}\|\in O(t^{-1-\epsilon})$, which proves (\ref{e4.10}) by the Cook argument. \qed

\subsection{Hawking effect I}\label{sec5.0}
\subsubsection{The limit state}
In the rest of the paper we denote by  $\beta= 2\pi\kappa^{-1}$ the inverse Hawking temperature.

Let us denote by $\omega^{0}_{\infty, \beta}$ the gauge-invariant quasi-free  thermal state on $\fA_{\infty}$ with covariance:
\[
\omega^{0}_{\infty, \beta}\left(\psi^{*}(f)\psi(g)\right)= (f| (\one + \e^{- \beta b^{0}_{\infty}})^{-1}g), \ f,g\in \ch_{\infty}.
\]
This state restricts to a quasi-free state on $\fA_{0}^{\rm l}$, still denoted by $\omega^{0}_{\infty, \beta}$.

We denote by  $\omega^{V}_{0, {\rm vac}}$ the gauge-invariant quasi-free vacuum state on $\fA_{0}$ with covariance:
\[
\omega^{V}_{0, {\rm vac}}(\psi^{*}(f)\psi(g))= (f| \one_{\rr^{+}}(b_{0}^{V})g), \ f,g\in \ch_{0}.
\]
The state $\omega^{V}_{0, {\rm vac}}\circ \gamma^{\rm r, int}$ is a gauge-invariant state on $\fA_{0}^{\rm r}$, which is even by Lemma \ref{4.8}. 

Since $\ch_{0}= \ch_{0}^{\rm l}\oplus \ch_{0}^{\rm r}$, we can by  Def. \ref{app1.6} define the following state on $\fA_{0}$:
\begin{definition}\label{defodef}
 We set
 \[
\omega_{\rm H, I}:= \omega^{0}_{\infty, \beta}\widehat{\otimes}\,\,(\omega^{V}_{0, {\rm vac}}\circ\gamma^{\rm r, int}), 
\]
which is a state on $\fA_{0}$.
\end{definition}
\subsubsection{Main result I}
The following theorem is the main result of this section.
\begin{theoreme}\label{5.0}
\[
\lim_{t\to +\infty}\omega^{V}_{0, {\rm vac}}\circ\tauvint(0,t)\circ \alpha^{t}(A)= \omega_{\rm H, I}(A), \ A\in \fA_{0}.
\] 
\end{theoreme}
\proof By linearity and density we may assume that $A= A_{1}\times A_{2}$, $A_{1}\in \fA_{0}^{\rm l}$, $A_{2}\in \fA_{0}^{\rm r}$.  By Prop. \ref{4.7} we have $\tauvint\circ \alpha^{t}(A_{2})= \gamma^{d, \rm int}(A_{2})+ o(t^{0})$. Applying Lemma \ref{5.1} we obtain that

\[
\lim_{t\to +\infty}\omega^{V}_{0, {\rm vac}}(A_{1}A_{2})=  \omega^{0}_{\infty, \beta}\widehat{\otimes}\,\,\omega^{V}_{0, {\rm vac}}\circ \gamma^{\rm r, int}(A_{1}A_{2}),
\]
which completes the proof of the theorem. \qed
\begin{lemma}
 \label{5.1}
 Let $A_{1}\in \fA^{\rm l}_{0}$, $A_{2}\in \fA_{0}$. Then
 \[
\lim_{t\to +\infty}\omvac(\tauvint(0,t)\circ \alpha^{t}(A_{1})A_{2})=  \omega^{0}_{\infty, \beta}\widehat{\otimes}\,\,\omvac(A_{1}A_{2}),.
\]
\end{lemma}
\proof By linearity and density we can assume that 
\[
A_{1}= \prod_{i=1}^{n_{1}} \psi^{*}(f_{i})\prod_{i=1}^{p_{1}}\psi(g_{i}),  A_{2}= \prod_{i=1}^{n_{2}} \psi^{*}(f_{n_{1}+i})\prod_{i=1}^{p_{2}}\psi(g_{p_{1}+i}),
\]
where
\[
\begin{array}{rl}
&f_{i}, g_{j}\in \ch_{0}^{\rm l}, \hbox{ for }1\leq i\leq n_{1}, \ 1\leq j \leq p_{1},\\[2mm]
&f_{n_{1}+i}, g_{p_{1}+j}\in \ch_{\infty},\hbox{ for } 1\leq i \leq n_{2}, \ 1\leq j\leq p_{2}.
\end{array}
\]
To simplify notation we set
\[
\gamma^{t}_{\rm int}:= \tauvint(0,t)\circ \alpha^{t}, \ \gamma^{t}:=\tauv(0,t)\circ \alpha^{t},
\]
so that from Prop. \ref{4.5} we have $\gamma^{t}_{\rm int}(A_{1})= \gamma^{t}(A_{1})+ o(t^{0})$. It follows that
\[
\gamma^{t}_{\rm int}(A_{1})=\prod_{i=1}^{n_{1}}\psi^{*}(u^{V}(0,t)f_{1i}^{t})\prod_{i=1}^{p_{1}}\psi(g_{1i}^{t})+ o(t^{0}).
\]
Using the CAR and Prop. \ref{2.6bis}, we obtain that:
\[
\begin{array}{rl}
&\gamma^{t}_{\rm int}(A_{1})A_{2}\\[2mm]
=&  (-1)^{n_{2}(n_{1}+ p_{1})}\prod_{i=1}^{n_{2}} \psi^{*}(f_{2i})\prod_{i=1}^{n_{1}}\psi^{*}(u^{V}(0,t)f_{1i}^{t})\prod_{i=1}^{p_{1}}\psi(g_{1i}^{t})\prod_{i=1}^{p_{2}}\psi(g_{2i})+o(t^{0}).
\end{array}
\]
Since $\omega^{V}_{0,\vac}$ is a gauge invariant quasi-free state (see Subsect. \ref{secapp1.3}), we see that $\omvac(\gamma^{t}_{\rm int}(A_{1})A_{2})=o(t^{0})$ if $n_{1}+ n_{2}\neq p_{1}+ p_{2}$, and if $n_{1}+n_{2}= p_{1}+ p_{2}=n$ we have:
\beq\label{e5.1}
\begin{array}{rl}
&\omvac(\gamma^{t}_{\rm int}(A_{1})A_{2})\\[2mm]
=&(-1)^{n_{2}(n_{1}+ p_{1})}\sum_{\sigma\in S_{n}}\epsilon(\sigma)\prod_{k=1}^{n} \omvac(\psi^{*}(F^{t}_{k})\psi(G_{\sigma(k)}^{t}))+o(t^{0}),
\end{array}
\eeq
where:
\[
F_{k}^{t}= \left\{\begin{array}{l}
u^{V}(0,t)f_{k}^{t}\hbox{ for }1\leq k\leq n_{1},\\
f_{k}\hbox{ for }n_{1}+1\leq k\leq n,
\end{array}\right., \ G_{k}^{t}= \left\{\begin{array}{l}
u^{V}(0,t)g_{k}^{t}\hbox{ for }1\leq k\leq p_{1},\\
g_{k}\hbox{ for }p_{1}+1\leq k\leq n.
\end{array}\right.
\]
Recall that $\omvac(\psi^{¨}(f)\psi(g))= (f| \one_{\rr^{+}}(b_{0}^{V})g)_{\ch_{0}}$ and $\wlim u^{V}(0,t)f^{t}=0$ for $f\in \ch^{\rm l}_{0}$ by Prop. \ref{2.6bis}. We see the sum on the r.h.s. is $o(t^{0})$ unless $n_{1}=p_{1}$ and $n_{2}=p_{2}$. If this is the case the only permutations $\sigma$ contributing to the sum are of the form $\sigma_{1}\times \sigma_{2}$ where $\sigma_{i}\in S_{n_{i}}$. Collecting these terms we obtain that:
\[
\omvac(\gamma^{t}_{\rm int}(A_{1})A_{2})= \omvac(\gamma^{t}(A_{1}))\omvac(A_{2}) + o(t^{0}).
\]
By the result of Bachelot \cite{Ba1} recalled in Thm. \ref{2.7} we know that 
\[
\lim_{t\to +\infty}\omvac(\gamma^{t}(A_{1}))= \omega^{0}_{\infty,\beta}(A_{1}).
\]
Now we use Remark \ref{rem1} and the definition of  the $\zz_{2}-$graded tensor product of two states (see Lemma \ref{app1.5}) to see that 
\[
\lim_{t\to +\infty}\omvac(\gamma^{t}_{\rm int}(A_{1})A_{2})= \omega^{0}_{\infty,\beta}\widetilde{\otimes}\omvac(A_{1}A_{2}),
\]
which completes the proof of the lemma. \qed

\subsection{Change of initial state}\label{sec5.1}
We assumed in Subsect. \ref{sec2.1} that the star was stationary in $t\leq 0$. It is hence natural to take as dynamics in the past the  stationary {\em interacting }dynamics  $\tauvint_{0}(s,t)$ on $\fA_{0}$ defined as follows: we first define the stationary analog of $\tau^{V}(s,t)$, acting on $\fA_{0}$ by
\[
\tau^{V}_{0}(s,t)\psi^{(*)}(f):= \psi^{(*)}(\e^{\i t b^{V}_{0}}f), \ f\in \ch_{0}.
\]
We can then define the stationary interacting dynamics $\tauvint_{0}(s,t)$ associated to $I$ in Def. \ref{4.1}. It suffices to repeat the construction in Subsect. \ref{sec4.1} with $\tau^{V}_{0}(s,t)$ instead of $\tau^{V}(s,t)$.

An adapted  choice of  the initial state in Thm. \ref{5.0} would be an even state $\tilde{\omega}$ 
on $\fA_{0}$, {\em invariant }under $\tauvint_{0}(s,t)$.   The following  easy result shows that Thm. \ref{5.0} will extend to $\tilde{\omega}$, provided that $\tilde{\omega}$ belongs to the  {\em folium of }$\omega^{V}_{0,{\rm vac}}$, i.e. is represented by a density matrix in the GNS representation of  $\omega^{V}_{0,{\rm vac}}$. Recall that such states are physically interpreted as local perturbations of $\omega^{V}_{0, {\rm vac}}$.
\begin{corollary}\label{5.2}
 Let $\tilde{\omega}$ a state on $\fA_{0}$ which is even and belongs to the folium of  $\omega^{V}_{0, {\rm vac}}$. Then 
 \[
\lim_{t+\infty}\tilde{\omega}\circ \tauvint(0,t)\circ \alpha^{t}(A)= \tilde{\omega}_{\rm H, I}(A),\ A\in \fA_{0},
\]
where:
\[
 \tilde{\omega}_{\rm H, I}= \omega^{0}_{\infty, \beta}\widehat{\otimes}\,\, (\tilde{\omega}\circ \gamma^{\rm r, int}). 
\]
\end{corollary}
\proof Since $\tilde{\omega}$ belongs to the folium of  $\omega^{V}_{0,{\rm vac}}$, we are, by linearity and density, reduced to compute the limit:
\[
\lim_{t\to +\infty}\omega^{V}_{0,{\rm vac}}(P^{*}(\psi, \psi^{*})\gamma^{t}(A)_{1}A_{2}P(\psi, \psi^{*})),
\]
where $A_{1}\in \fA^{\rm l}_{0}$, $A_{2}\in \fA_{0}$ and $P(\psi, \psi^{*})$ is a polynomial in $\CAR_{\rm alg}(\ch_{0})$. Moreover since $\tilde{\omega}$ is even, we see that $P(\psi, \psi^{*})\in \CAR_{{\rm alg}, 0}(\ch_{0})$. By the same argument as in the proof of Lemma \ref{5.1}, we see that 
\[
P^{*}(\psi, \psi^{*})\gamma^{t}(A)_{1}A_{2}P(\psi, \psi^{*})= \gamma^{t}(A)_{1}P^{*}(\psi, \psi^{*})A_{2}P(\psi, \psi^{*})+ o(t^{0}),
\]
hence as in Lemma \ref{5.1} we have:
\[
\begin{array}{rl}
&\lim_{t\to+\infty}\omega^{V}_{0,{\rm vac}}(P^{*}(\psi, \psi^{*})\gamma^{t}(A)_{1}A_{2}P(\psi, \psi^{*}))\\[2mm]
=& \omega^{0}_{\beta}(A_{1}) \omega^{V}_{0, {\rm vac}}(P^{*}(\psi, \psi^{*})A_{2}P(\psi, \psi^{*}))=  \omega^{0}_{\beta}(A_{1})\tilde{\omega}(A_{2}).

\end{array}
\]
We can then complete the proof as in Thm. \ref{5.0}. \qed

\subsection{Existence of interacting initial vacua}\label{sec5.2}
It remains to construct even states $\tilde{\omega}$ which belong to the folium of $\omega^{V}_{0,{\rm vac}}$ and  are invariant under $\tauvint_{0}(s,t)$. To do this it is convenient to work in the GNS representation of  the vacuum state $\omega^{V}_{0, {\rm vac}}$, i.e.  the {\em Fock representation}.
We refer the reader to Subsect. \ref{secapp.3bis}.

 Recall that $b^{V}_{0}$ is defined in Subsect. \ref{diri}. It is easy to show that
\begin{equation}
\label{e5.1bis}
\sigma_{\rm ess}(b^{V}_{0})= ]-\infty, -m]\cup [m,+\infty[.
\end{equation}
We assume that ${\rm Ker}b^{V}_{0}= \{0\}$ and  equip $\ch_{0}$ with the complex structure $\ii:= \i \, {\rm sgn}(b^{V}_{0})$, and denote   by $\cZ$ the associated one-particle space. If $\pi_{F}$ is the corresponding Fock representation we have:
\[
\omega^{V}_{0, {\rm vac}}(A)= (\Omega| \pi_{F}(A)\Omega),
\]
where $\Omega\in \Gamma_{\rm a}(\cZ)$ is the vacuum vector.  
 In other words $(\Gamma_{\rm a}(\cZ), \pi_{F}, \Omega)$ is the GNS triple associated to   $\omega^{V}_{0, {\rm vac}}$. 

From Subsect. \ref{secapp.3bis} we know that:
 \[
\pi_{F}(\tau^{V}_{0}(s,t)A)= \e^{\i (s-t) H_{0}}\pi_{F}(A)\e^{\i (t-s)H_{0}}, \ A\in \fA_{0},
\]
for $H_{0}= \d\Gamma(| b^{V}_{0}|)$, and 
 if $Q= \d\Gamma({\rm sgn}(b^{V}_{0}))$, then:
\[
\psi^{(*)}_{F}(\e^{\ i \theta}f)= \e^{\i \theta Q}\psi^{(*)}_{F}(f)\e^{-\i \theta Q}, \ f\in \ch_{0}, \ \theta\in \rr.
\]
 It is also well known  that if
\[
H:=H_{0}+ \pi_{F}(I),
\]
then
\[
\pi_{F}(\tauvint_{0}(s,t)(A))= \e^{\i (s-t)H}\pi_{F}(A)\e^{\i(t-s)H}, \ A\in \fA_{0}.
\]
 Since $\tauvint_{0}(s,t)$ is implemented by $\e^{\i (s-t)H}$ in the Fock representation, eigenvectors of $H$ will yield invariant states for $\tauvint_{0}(s,t)$, which  obviously belong to the folium of $\omega^{V}_{0, {\rm vac}}$.
 
   Existence of eigenvectors is ensured by the following theorem, whose proof follows by adapting arguments in \cite{A,DG}.
\begin{theoreme}\label{5.3}
On has
\[
 \sigma_{\rm ess}(H)= [\inf \sigma(H)+ m, +\infty[.
\]
Therefore $\inf \sigma(H)$ is an eigenvalue of $H$.
\end{theoreme}
To be able to apply Corollary \ref{5.2}, we need however the existence of an {\em even} eigenstate $\psi$ of $H$, i.e.  such that $Q\psi= 2n\psi$ for some $n\in \zz$.  Note that since $I$ is even we have $[H, Q]=0$, which does not imply the existence of even eigenstates of $H$. However this is clearly true for small interactions. In fact setting $H(\lambda)= H_{0}+ \lambda I$ and $E(\lambda)= \inf \sigma(H(\lambda))$ we have
\begin{lemma}
 Assume $|\lambda|$ is small enough.Then $\one_{\{E(\lambda)\}}(H(\lambda))$ is rank one and $Q\one_{\{E(\lambda)\}}(H(\lambda))=0$.
\end{lemma}
Therefore for $\lambda$ small enough, $H(\lambda)$ has a unique ground state $\Omega(\lambda)$ of zero charge and the associated state satisfies the hypotheses of Corollary \ref{5.2}.

\section{Hawking effect II}\label{sec6}\init
In this section we study the Hawking effect in  case II (see Subsect. \ref{sec1.3}).  Compared to Sect. \ref{sec5}, the observable is not translated to the left, therefore the influence of the potential $V$ and of the non-linear self-interaction has to be taken into account. To this end we use tools from scattering theory, both for classical and quantum dynamics.
\subsection{Asymptotic velocity for Dirac equations}\label{sec6.1}
In this subsection we state some results of Daud\'e \cite{Da} on the existence of the {\em asymptotic velocity observable} for stationary Dirac equations. The asymptotic velocity provides a convenient way to separate left and right propagating initial states. More details can be found in \cite{Da}.

\begin{theoreme}\label{6.1}
 Let $\chi\in \coinf(\rr, \cc^{2})$. Then
 \ben
\item the limits
 \[
\chi^{\pm}:= \slim_{t\to \pm\infty}\e^{-\i t b^{V}_{\infty}}\chi(\xt)\e^{\i t b^{V}_{\infty}}\hbox{ exist.}
\]
\item there exist  bounded selfadjoint operators $P^{\pm}$ on $\ch_{\infty}$ such that $\chi^{\pm}= \chi(P^{\pm})$ for $\chi\in \coinf(\rr, \cc^{2})$.
\item one has
\[
[P^{\pm}, b^{V}_{\infty}]=0, \ \sigma(P^{\pm})= [-1, 1],  \ \one_{\{0\}}(P^{\pm})= \one_{\rm pp}(b^{V}_{\infty}).
\]
\een
\end{theoreme}
\begin{remark}\label{6.2}
 Since it is known (see e.g. \cite[Lemma III.1]{Ba1} that $b^{V}_{\infty}$ has no eigenvalues, we have actually $\one_{\{0\}}(P^{\pm})=0$, i.e. any initial state has a non-vanishing asymptotic velocity.
\end{remark}
We will only use the future asymptotic velocity $P^{+}$ which we will denote simply by $P$. Moreover we will set
\[
P^{{\rm l/r}}:= \one_{\rr^{\mp}}(P), 
\]
so that by Remark \ref{6.2} we have
\[
P^{\rm l}+ P^{\rm r}= \one.
\]
We set now $V^{\rm l/r}:=\lim_{x\to \mp\infty}V(x)$, so that $V^{\rm l}=0$, $V^{\rm r}= V_{\infty}$, see Subsect. \ref{sec2.1}. We set also
\[
b_{\infty}^{\rm l/r}:= L D_{x}+ V^{\rm l/r}, \hbox{ with domain }H^{1}(\rr, \cc^{2}), \hbox{ acting on }\ch_{\infty}.
\]
From Thm. \ref{6.1} and the short-range nature of the interaction $V$ (see (\ref{e.hyp1})), we obtain by standard arguments the existence of wave operators:
\def\gd{{\rm l/r}}
\begin{proposition}
 \label{6.2bis}
 The limit
 \[
\slim_{t\to +\infty}\e^{-\i t b_{\infty}^{\rm l}}\e^{\i t b_{\infty}^{V}}=: w_{\infty}^{\rm l}
\]
exists on $P^{\rm l}\ch_{\infty}$. 
\end{proposition}

Prop. \ref{6.2bis} yields the following result for free quantum dynamics.
\begin{proposition}
 \label{6.3}
 The limit
 \[
 \gamma_{\infty}^{\rm l}:=\slim_{t\to +\infty} \tau_{\infty}^{V^{\rm l}}(t,0)\circ \tau^{V}_{\infty}(0,t)\hbox{ exists on }\CAR(P^{\rm l}\ch_{\infty})
\]
The map $\gamma_{\infty}^{\rm l}: \CAR(P^{\rm l}\ch_{\infty})\to \CAR(\ch_{\infty})= \fA_{\infty}$ are $*-$morphisms with
\[
\gamma_{\infty}^{\rm l}(\psi^{(*)}(f))= \psi^{(*)}(w^{\rm l}f), \ f\in P^{\rm l}\ch_{\infty}.
\]
\end{proposition}
The similar limit $\gamma_{\infty}^{\rm r}$ with $V^{\rm l}$ replaced by $V^{\rm r}$ and $P^{\rm l}$ replaced by $P^{\rm r}$ exists on $\CAR(P^{\rm r}\ch_{\infty})$ also exists, but will play no role in the sequel. The 'right' analog of $\gamma^{\rm l}_{\infty}$ is the wave morphism $\gamma^{\rm r}_{0}$ introduced below in Prop. \ref{6.3c}.
\subsection{Wave morphisms}\label{sec6.2}
We now prove an analog of Prop. \ref{6.3} for interacting dynamics.
\begin{theoreme}
 \label{6.4}
 The limit
 \[
\gamma_{\infty}^{\rm int}:= \slim_{t\to +\infty}\tauvint_{\infty}(t,0)\circ \tau^{V}_{\infty}(0,t)\hbox{ exists on }\fA_{\infty}
\]
and is a $*-$morphism of $\fA_{\infty}$.
\end{theoreme}
The morphism $\gamma_{\infty}^{\rm int}$ is an example of a {\em wave morphism}.

{\bf Proof of Thm. \ref{6.4}.} The proof relies once again on the Cook argument, combined with minimal velocity estimates for the Dirac equation. For more details on minimal velocity estimates see \cite{Da}.  Since $b^{V}_{\infty}$ has no point spectrum we see that the space $\cD$ of vectors in $\cap_{n\in \nn}\Dom \langle x\rangle^{n}$ such that $f= \chi(b^{V}_{\infty})f$ for some $\chi\in \coinf(\rr\backslash [-m, m])$ is dense in $\ch_{\infty}$.
The strong minimal velocity estimates (see \cite{Da}) give
\begin{equation}
\label{e6.1}
\forall f \in\cD, \ \exists \  0<c_{0}<1 \hbox{ such that }\| \one_{[0, c_{0}]}(\frac{|x|}{t})\e^{\i t b^{V}_{\infty}}f\|\in O(t^{-N}), \ \forall \ N\in \nn.
\end{equation}
We can now argue as in the proof of Prop. \ref{4.7}, since by (\ref{e6.1}) $t\mapsto (g|\e^{\i t b^{V}_{\infty}}f)$ is integrable for $f\in \cD$. \qed

 We denote by $E_{t}: \fA_{\infty}\to \fA_{t}$ the $*-$homomorphism defined in  \ref{astolito}, associated to the inclusion $\ch_{t}\subset \ch_{\infty}$. Since $\cup_{t\geq 0}\ch_{t}$ is dense in $\ch_{\infty}$, we have:
 \begin{equation}
\label{astili}
\slim_{t\to +\infty} E_{t}= \one, \hbox{ in }\fA_{\infty}.
\end{equation}

We now combine Thm. \ref{6.4} with Lemma \ref{4.4} to obtain the following result.

\begin{proposition}
 \label{6.5}
Let $A= \gamma^{\rm int}_{\infty}(B)\in \fA_{\infty}$. Then for any $\epsilon>0$ there exist $C_{\epsilon}, T_{\epsilon}>0$ such that
\[
\sup_{t/2+ C_{\epsilon}\leq s \leq t/2 + 2C_{\epsilon}, \ t\geq T_{\epsilon}}\| \tauvint(s,t)\circ E_{t}(A)- \tau^{V}_{\infty}(s,t)(B)\| \leq \epsilon.
\]
\end{proposition}
\begin{remark}\label{igorito}
We do not know if  $A\in\gamma^{\rm int}_{\infty}\fA_{\infty}$ belongs to $\fA_{t}$ for all $t$ large enough, so  a priori $\tauvint(s,t)(A)$ does not makes sense. Replacing $A$  by $E_{t}(A)\in \fA_{t}$ fixes this problem, at the price of an  error $\|A- E_{t}(A)\|$ which is $o(t^{0})$. \end{remark}
\proof Since  $A= \gamma^{\rm int}_{\infty}(B)$ we have:
\[
\tau^{V}_{\infty}(0,t)(B)= \tauvint_{\infty}(0,t)(A)+ o(t^{0}), \ t\to +\infty.
\]
Since $\tau^{V}_{\infty}$ and $\tauvint_{\infty}$ are stationary dynamics, this implies that for any $c>0$:
\beq\label{e6.2}
\lim_{t\to +\infty}\sup_{0\leq s\leq t/2 +c}\|\tau^{V}_{\infty}(s, t)(B)- \tauvint_{\infty}(s,t)(A)\| =0.
\eeq
Since $\bigcup_{J\Subset \rr}\fA(J)$ is dense in $\fA_{\infty}$, we can for any $\epsilon>0$ find $J_{\epsilon}\Subset \rr$ and $A_{\epsilon}\in\fA(J_{\epsilon})$ such that $\|A- A_{\epsilon}\|\leq \epsilon/8$, hence:
\begin{equation}
\label{tar1}
\sup_{s,t}\| \tauvint_{\infty}(s,t)(A)- \tauvint_{\infty}(s,t)(A_{\epsilon})\|\leq \epsilon/8.
\end{equation}
By Lemma \ref{4.4} there exists $C_{\epsilon}= C(J_{\epsilon})$ such that:
\beq\label{tar3}
\tauvint(s,t)(A_{\epsilon})= \tauvint_{\infty}(s,t)(A_{\epsilon}), \ \forall \ t/2+C_{\epsilon}\leq s\leq t.
\eeq
Next by (\ref{astili}), we can find $T_{\epsilon}$ such that $\sup_{t\geq T_{\epsilon}}\| A- E_{t}(A)\|\leq \epsilon/8$, hence 
\beq\label{tar2}
\sup_{s\leq t, t\geq T_{\epsilon}}\| \tauvint(s,t)(A_{\epsilon})-\tauvint(s,t)\circ E_{t}(A)\|\leq \epsilon/4.
\eeq
Combining (\ref{tar1}), (\ref{tar3}) and (\ref{tar2}) we obtain:
\[
\sup_{t/2+C_{\epsilon}\leq s\leq t, \ t\geq T_{\epsilon}}\|\tauvint(s,t)\circ E_{t}(A)- \tauvint_{\infty}(s,t)(A)\|\leq \epsilon/2.
\]
Using also (\ref{e6.2}) we obtain the proposition. \qed
\medskip

%We now state an simpler version of Prop. \ref{6.5}, assuming the asymptotic completeness of $\gamma^{\rm int}_{\infty}$.
%\begin{proposition}\label{6.5bis}
%Assume additionally that $\gamma^{\rm int}_{\infty}\fA_{\infty}= \fA_{\infty}$ and let $A\in \fA_{J}$ for some $J\Subset \rr$. Then for any $\epsilon>0$ there exist $C_{\epsilon}, T_{\epsilon}>0$ such that
%\[
%\sup_{t/2+ C_{\epsilon}\leq s \leq t/2 + 2C_{\epsilon}, \ t\geq T_{\epsilon}}\| \tauvint(s,t)(A)- \tau^{V}_{\infty}(s,t)\circ (\gamma^{\rm int}_{\infty})^{-1}(A)\| \leq \epsilon.
%\]
%\end{proposition}
%\proof the proof is similar to that of Prop. \ref{6.5}. We do not need to introduce $E_{t}(A)$ and can replace $A_{\epsilon}$ in the proof of Prop. \ref{6.5} by $A$. Details are left to the reader. \qed

\subsection{Left propagation}\label{sec6.3}
Recall that the subspaces $\ch^{\gd}_{t}$ were defined in (\ref{e2.2}).
\begin{lemma}
 \label{6.6}
 We have $\Ran w^{\rm l}= \ch^{\rm l}_{\infty}$. It follows that $\gamma^{\rm l}_{\infty}: \CAR(P^{\rm l}\ch_{\infty})\ \tilde{\to} \ \CAR(\ch_{\infty}^{\rm l})$ is a $*-$isomorphism. 
\end{lemma}
\proof 
 Let us denote by $\tilde{P}$ the asymptotic velocity for $b^{\rm l}_{\infty}$. Since $V^{\rm l}=0$ we have $b^{\rm l}_{\infty}= -LD_{x}$. If $x(t)= \e^{-\i tb^{\rm l}_{\infty}}x\e^{\i t b^{V}_{\infty}}$, we see that $x(t)= -tL +x$, from which it follows that $\tilde{P}=-L$.
 It is well known (see e.g. \cite{Da}) that the wave operator $w^{\rm l}$ intertwines $P$ and $\tilde{P}$, hence 
\[
\Ran w^{\rm l}= \Ran\one_{\rr^{-}}(\tilde{P})= \Ran\one_{\rr^{+}}(L)= \ch^{\rm l}_{\infty}.
\]
This completes the proof of the lemma. \qed

\begin{proposition}\label{6.7}
 Let $A= \gamma^{\rm int}_{\infty}(B^{\rm l})$, $ B^{\rm l}\in \CAR(P^{\rm l}\ch_{\infty})$. Then for any $\epsilon>0$ there exist $C_{\epsilon}, T_{\epsilon}>0$ such that
 \[
\sup_{t\geq T_{\epsilon}}\|  \tauvint(t/2+C_{\epsilon}, t)\circ E_{t}(A)- \alpha^{t/2- C_{\epsilon}}\circ \gamma^{\rm l}_{\infty}(B^{\rm l})\|\leq \epsilon.
\]
\end{proposition}
\proof 
By Prop. \ref{6.5} there exist $C_{\epsilon}, \tilde T_{\epsilon}$ such that
\begin{equation}
\label{e6.01}
\sup_{t\geq \tilde T_{\epsilon}}\| \tauvint(t/2+ C_{\epsilon}, t)\circ E_{t}(A)- \tau^{V}_{\infty}(t/2+C_{\epsilon},t)(B^{\rm l})\|\leq \epsilon/2.
\end{equation}
Moreover since $B^{\rm l}\in \CAR(P^{\rm l}\ch_{\infty})$, we have
\[
\lim_{s\to +\infty} \tau^{V}_{\infty}(0,s)(B^{\rm l}) - \tau^{0}_{\infty}(0, \sigma)\circ \gamma^{\rm l}_{\infty}(B^{\rm l})=0,
\]
by Lemma \ref{6.6}. We now note that $\tau^{V}_{\infty}, \tau^{0}_{\infty}$ are stationary, $\gamma^{\rm l}_{\infty}(B^{\rm l})\in \CAR(\ch^{\rm l}_{\infty})$, and $\tau^{0}_{\infty}(0, s)= \alpha^{s}$ on $\CAR(\ch^{\rm l}_{\infty})$, since $u^{0}_{\infty}(0,s)f= f^{s}$ for $s\in\ch^{\rm l}_{\infty}$, by (\ref{e2.3}). It follows that we can find $T_{\epsilon}\geq \tilde{T_{\epsilon}}$ such that
\[
\|  \tauvint(t/2+C_{\epsilon}, t)\circ E_{t}(A)- \alpha^{t/2- C_{\epsilon}}\circ \gamma^{\rm l}_{\infty}(B^{\rm l})\|\leq \epsilon,
\]
for $t\geq T_{\epsilon}$. This completes the proof of the proposition.
 \qed

\subsection{Right propagation}\label{sec6.4}
The following lemma means that for an initial observable propagating to the right, the influence of the boundary condition on the star can be forgotten.
\begin{lemma}
 \label{6.8}
 For any $c>0$ and $B^{\rm r}\in \CAR(P^{\rm r}\ch_{\infty})$ one has:
 \[
\tau^{V}(0, t/2+c)\circ \tau^{V}_{\infty}(t/2+c,t) (B^{\rm r})= \tau^{V}_{\infty}(0,t)(B^{\rm r})+ o(t^{0}), \ t\to +\infty.
\]
\end{lemma}
\proof By the usual arguments of linearity and density it suffices to prove that
\begin{equation}
\label{e6.3}
u^{V}(0, t/2+c)\circ u^{V}_{\infty}(t/2+c,t)f= u^{V}_{\infty}(0,t)f+ o(t^{0}), \ f\in  \Dom b^{V}_{\infty}\cap P^{\rm r}\ch_{\infty}, \ t\to +\infty.
\end{equation}
Using that  $u^{V}_{\infty}(s,t)= \e^{\i (s-t)b^{V}_{\infty}}$ we can moreover assume that there exists $\epsilon>0$ such that:
\beq\label{e6.4}
\begin{array}{rl}
&u^{V}_{\infty}(0,t)f= \chi(\frac{x}{t}\geq \epsilon)u^{V}_{\infty}(0,t)f+ o(t^{0}),\\[2mm]
&u^{V}_{\infty}(t/2+c, t)f= \chi(\frac{x}{t}\geq \epsilon)u^{V}_{\infty}(t/2+c, t)f+ o(t^{0}).
\end{array}
\eeq
To simplify notation let us set
\[
f_{s,t}= \chi(\frac{x}{t})u^{V}_{\infty}(s,t)f.
\]
From (\ref{e6.4}) we obtain
\[
\begin{array}{rl}
&\| u^{V}(0, t/2+c)\circ u^{V}_{\infty}(t/2+c, t)f- u^{V}_{\infty}(0,t)f\|\\[2mm]
=&\| u^{V}(0, t/2+c)f_{t/2+c, t}- f_{0,t}\|+ o(t^{0})\\[2mm]
=&\| f_{t/2+c, t}- u^{V}(t/2+c, 0)f_{0,t}\| + o(t^{0})\\[2mm]
\leq & \| \int_{0}^{t/2+c} \p_{s}u^{V}(t/2+c, s)f_{s,t}\| ds + o(t^{0}).
\end{array}
\]
We have
\[
\p_{s}u^{V}(t/2+c, s)f_{s,t}=  -\i u^{V}(t/2+c, s) \left(  b^{V}_{s} \chi(\frac{x}{t}\geq \epsilon) -  \chi(\frac{x}{t}\geq \epsilon)b^{V}_{\infty}\right)u^{V}_{\infty}(s,t)f.
\]
Since $f\in\Dom b^{V}_{\infty}= H^{1}(\rr,, \cc^{2})$, we know that $f_{s,t}\in H^{1}(\rr, \cc^{2})$. Since moreover $f_{s,t}$ is equal to $0$ in $x\leq \epsilon t$, we have $f_{s,t}\in \Dom b^{V}_{s}$ for $0\leq s\leq t/2 +c$ and 
\[
\begin{array}{rl}
& \left(  b^{V}_{s} \chi(\frac{x}{t}\geq \epsilon) -  \chi(\frac{x}{t}\geq \epsilon)b^{V}_{\infty}\right)u^{V}_{\infty}(s,t)f\\[2mm]
=&[b^{V}_{\infty}, \chi(\frac{x}{t}\geq \epsilon)] u^{V}_{\infty}(0,t)f=  \frac{1}{t}L \chi'(\frac{x}{t}\geq \epsilon)u^{V}_{\infty}(0,t)f.
\end{array}
\]
It follows that
\beq\label{e6.5}
\begin{array}{rl}
& \| \int_{0}^{t/2+c} \p_{s}u^{V}(t/2+c, s)f_{s,t}\| ds\leq \frac{1}{t} \int_{0}^{t/2+c} \| \chi(\frac{x}{t}\simeq \epsilon)u^{V}_{\infty}(s,t)f\| ds\\[2mm]
\leq&\frac{1}{t} \int_{t/2-c}^{t}\| \chi(\frac{x}{t}\simeq \epsilon)\e^{- \i \tau b^{V}_{\infty}}f\| d\tau,
\end{array}
\eeq
setting $\tau=t-s$.   Denoting by $R_{t}(f)$ the r.h.s. in (\ref{e6.5}) we have $\|R_{t}(f)\|\leq C \| f\|$, $f\in \ch_{\infty}$ and from (\ref{e6.1}) $R_{t}(f)\in O(t^{-\infty})$ for $f\in \cD$ hence $R_{t}(f)\in o(t^{0})$ for any $f\in \ch_{\infty}$. This proves (\ref{e6.3}) and completes the proof of the lemma. \qed

\begin{proposition}
 \label{6.9}
 Let $A= \gamma^{\rm int}_{\infty}(B^{\rm r})$, $B^{\rm r}\in \CAR(P^{\rm r}\ch_{\infty})$. Then for any $\epsilon>0$ there exists $T_{\epsilon}>0$ such that
 \[
\sup_{t\geq T_{\epsilon}}\|  \tauvint(0,t)\circ E_{t}(A)- \tau^{V}_{\infty}(0, t)(B^{\rm r})\|\leq \epsilon.
\]
 \end{proposition}
\proof Let $\epsilon>0$. By Prop. \ref{6.5} there exist $C_{\epsilon}, \tilde{T}_{\epsilon}$ such that
\beq\label{dito}
\begin{array}{rl}
\tauvint(0,t)\circ E_{t}(A)=& \tauvint(0, t/2+ C_{\epsilon})\circ \tauvint(t/2+ C_{\epsilon}, t)\circ E_{t}(A)\\[2mm]
=& \tauvint(0, t/2+ C_{\epsilon})\circ \tau^{V}_{\infty}(t/2+ C_{\epsilon}, t)(B^{\rm r})+ R_{\epsilon}(t),
\end{array}
\eeq
where $\|R_{\epsilon}(t)\|\leq \epsilon/4$ for $t\geq \tilde{T}_{\epsilon}$. Recall that the dense subspace $\cD\subset \ch_{\infty}$ was introduced in the proof of Thm. \ref{6.4}. Since $B^{\rm r}\in \CAR(P^{\rm r}\ch_{\infty})$, we can by density find $\tilde{B}^{\rm r}\in \CAR_{\rm alg}(\cD\cap P^{\rm r}\ch_{\infty})$ such that $\|B^{\rm r}- \tilde{B}^{\rm r}\|\leq \epsilon/4$. It follows then from (\ref{dito}) that
\beq\label{diti}
\sup_{t\geq \tilde{T}_{\epsilon}}\|\tauvint(0,t)\circ E_{t}(A)- \tauvint(0, t/2 + C_{\epsilon})\circ \tau^{V}_{\infty}(t/2+ C_{\epsilon}, t)(\tilde{B}^{\rm r})\| \leq \epsilon/2.
\eeq
Next we have
\[
\begin{array}{rl}
&\tauvint(0, t/2 + C_{\epsilon})\circ \tau^{V}_{\infty}(t/2+ C_{\epsilon}, t)(\tilde{B}^{\rm r})\\[2mm]
=&R(0, t/2+ C_{\epsilon}) \tau^{V}_{\infty}(0,t)(\tilde{B}^{\rm r})R(0, t/2+ C_{\epsilon})^{*}+ o(t^{0}),
\end{array}
\]
by Lemma \ref{6.8} and the definition of $\tauvint$ in Def. \ref{4.2}. It follows that
\[
\begin{array}{rl}
&\tauvint(0, t/2 + C_{\epsilon})\circ \tau^{V}_{\infty}(t/2+ C_{\epsilon}, t)(\tilde{B}^{\rm r})\\[2mm]
=& \tau^{V}_{\infty}(0,t)(\tilde{B}^{\rm r})-\int_{0}^{t/2+ C_{\epsilon}}\p_{s}\left(R(s, t/2+ C_{\epsilon}) \tau^{V}_{\infty}(0,t)(\tilde{B}^{\rm r})R(s, t/2+ C_{\epsilon})^{*}\right)ds.
\end{array}
\]
Now
\[
\begin{array}{rl}
&\p_{s}\left(R(s, t/2+ C_{\epsilon}) \tau^{V}_{\infty}(0,t)(\tilde{B}^{\rm r})R(s, t/2+ C_{\epsilon})^{*}\right)\\[2mm]
=&\i R(s, t/2+ C_{\epsilon})[I(s,t/2+C_{\epsilon}),  \tau^{V}_{\infty}(0,t)(\tilde{B}^{\rm r})]R(s, t/2+ C_{\epsilon})^{*}.
\end{array}
\]
It remains to bound the norm of $[I(s,t/2+C_{\epsilon}),  \tau^{V}_{\infty}(0,t)(\tilde{B}^{\rm r})]$, since $R(\cdot, \cdot)$ is unitary.  
This amounts again to estimate scalar products of the form
\[
(u^{V}(s, t/2+C_{\epsilon})g| u^{V}_{\infty}(0,t)f),
\]
for $g$ compactly supported, $f\in\cD\cap P^{\rm r}\ch_{\infty}$.  We know that $u^{V}(s, t/2+ C_{\epsilon})g$ is supported in $\{|x|\leq t/2 + C_{\epsilon}+ C_{0}\}$ for $s\leq t/2+ C_{\epsilon}$ and $C_{0}>0$, since $g$ has compact support.  On the other hand if $f\in \cD\cap P^{\rm r}\ch_{\infty}$ we have by (\ref{e6.1})
\[
\|\one_{[0, c_{0}]}(\frac{|x|}{t}) u^{V}_{\infty}(0,t)f\|\in O(t^{-N}), \ \forall \ N\in\nn.
\]
It follows that 
\[
\begin{array}{rl}
&\|\tauvint(0, t/2 + C_{\epsilon})\circ \tau^{V}_{\infty}(t/2+ C_{\epsilon}, t)(\tilde{B}^{\rm r})- \tau^{V}_{\infty}(0,t)(\tilde{B}^{\rm r})\|\\[2mm]
\leq &  D_{\epsilon}| t/2+ C_{\epsilon}| \langle t\rangle^{-N},
\end{array}
\]
hence by (\ref{diti}) there exists $T_{\epsilon}\geq \tilde{T}_{\epsilon}$ such that
\[
\sup_{t\geq T_{\epsilon}}\| \tauvint(0,t)\circ E_{t}(A)- \tau^{V}_{\infty}(0, t)(\tilde{B}^{\rm r})\|\leq 3\epsilon/4.
\]
Since $\| B^{\rm r}- \tilde{B}^{\rm r}\|\leq \epsilon/4$, this  completes the proof of the proposition. \qed

We conclude this subsection by stating two easy scattering results for free dynamics with boundary conditions.
\begin{lemma}
 \label{6.3b}
 The limit
 \[
\slim_{t\to +\infty}\e^{\i t b_{0}^{V}}\e^{- \i t b^{V}_{\infty}}\hbox{ exists on }P^{\rm r}\ch_{\infty}.
\]
Moreover the above limit is unitary from $P^{\rm r}\ch_{\infty}$ to $\ch_{0}$.
\end{lemma}
\proof the proof follows by standard arguments (note that $\e^{-\i t b^{V}_{\infty}}f$ propagates to the right when $f\in P^{\rm r}\ch_{\infty}$, hence the boundary condition at $x= z(0)$ is irrelevant). \qed

Lemma \ref{6.3b} immediately implies the following proposition:
\begin{proposition}
 \label{6.3c}
 The limit
 \[
\gamma^{\rm r}_{0}:=\slim_{t\to +\infty} \tau^{V}_{0}(t,0)\circ \tau^{V}_{\infty}(0,t)\hbox{ exists on }\CAR(P^{\rm r}\ch_{\infty}),
\]
and $\gamma^{\rm r}_{0}: \CAR(P^{\rm r}\ch_{\infty})\tilde{\to}\CAR(\ch_{0})$ is a $*-$isomorphism.
\end{proposition}
Combining Props. \ref{6.9} and \ref{6.3c} we obtain the following proposition, which is the main result of this subsection.
\begin{proposition}
 \label{6.12}
 Let $A= \gamma^{\rm int}_{\infty}(B^{\rm r})$, $B^{\rm r}\in \CAR(P^{\rm r}\ch_{\infty})$. The for any $\epsilon>0$ there exists $T_{\epsilon}>0$ such that
 \[
\sup_{t\geq T_{\epsilon}} \| \tauvint(0,t)\circ E_{t}(A)- \tau^{V}_{0}(0,t)\circ \gamma^{\rm r}_{0}(B^{\rm r})\|\leq \epsilon. 
\]
\end{proposition}

\subsection{Hawking effect II}\label{sec6.5}
\subsubsection{The limit state}
Before stating our main result on the Hawking effect, let us introduce some notation. Recall that $\gamma^{\rm l}_{\infty}: \CAR(P^{\rm l}\ch_{\infty})\tilde{\to}\CAR(\ch^{\rm l}_{\infty})$ defined in Prop. \ref{6.3} is a $*-$isomorphism.  Similarly $\gamma^{\rm r}_{0}: \CAR(P^{\rm r}\ch_{\infty})\tilde{\to}\CAR(\ch_{0})$ defined in Prop. \ref{6.3c} is a $*-$isomorphism.

\begin{lemma}
Let us denote by  $\omega^{\rm l}_{\infty, \beta}$ the state on $\CAR(P^{\rm l}\ch_{\infty})$ equal to
 \[
\omega^{\rm l}_{\infty, \beta}:= \omega^{0}_{\infty, \beta}\circ \gamma^{\rm l}_{\infty},
\]
and by $\omega^{\rm r}_{\infty, {\rm vac}}$ the state on $\CAR(P^{\rm r}\ch_{\infty})$ equal to
  \[
\omega^{\rm r}_{\infty, {\rm vac}}:= \omega^{V}_{0, {\rm vac}}\circ \gamma^{\rm r}_{0}.
\]
Then 
\ben
\item $\omega^{\rm l}_{\infty, \beta}$ is  the restriction to $\CAR(P^{\rm l}\ch_{\infty})$ of the quasi-free thermal state on $\CAR(\ch_{\infty})$ with covariance
\[
(f| (\one + \e^{-\beta b^{V}_{\infty}})^{-1}f), \ f\in \ch_{\infty}.
\] 
\item $\omega^{\rm r}_{\infty, {\rm vac}}$ is the restriction to $\CAR(P^{\rm r}\ch_{\infty})$ of the quasi-free vacuum  state on $\CAR(\ch_{\infty})$ with covariance
\[
(f| \one_{\rr^{+}}(b^{V}_{\infty})f), \ f\in \ch_{\infty}.
\] 
\een
\end{lemma}
\proof 
(1) follows from the fact that $\gamma^{\rm l}_{\infty}$ is implemented by the wave operator  $w^{\rm l}$ defined in Prop. \ref{6.2bis}, which  intertwines $b_{\infty}^{V}$ and $b^{0}_{\infty}$. Similarly (2) follows from the intertwining properties of the wave operator constructed in Lemma \ref{6.3b}. \qed

Note that $\omega^{\rm l}_{\infty, \beta}$ and $\omega^{\rm r}_{\infty, {\rm vac}}$ are even states. Since $\ch_{\infty}= P^{\rm l}\ch_{\infty}\oplus P^{\rm r}\ch_{\infty}$,  we can  define the following state, acting on $\gamma^{\rm int}_{\infty}\CAR(\ch_{\infty})$:
\begin{definition}
We set:
 \[
 \omega_{\rm H, II}:= (\omega^{\rm l}_{\infty, \beta}\widehat{\otimes}\,\,\omega^{\rm r}_{\infty, {\rm vac}})\circ (\gamma^{\rm int}_{\infty})^{-1},
\]
which is a state on  $\gamma^{\rm int}_{\infty}\CAR(\ch_{\infty})$.
\end{definition}
\begin{remark}\label{ac}
  Note that the limit state $\omega_{\rm H, II}$ is a priori only defined on the $C^{*}$-algebra $\gamma^{\rm int}_{\infty}\CAR(\ch_{\infty})$ and not on the whole of $\CAR(\ch_{\infty})$. One can of course assume the  {\em asymptotic completeness} of  the wave morphism $\gamma^{\rm int}_{\infty}$:
  \[
{\rm (AC)} \ \gamma_{\infty}^{\rm int}\fA_{\infty}= \fA_{\infty},
\]
in which case the statement of Thm. \ref{6.13} below simplifies.

To our knowledge, the asymptotic completeness is  an essentially open question in the algebraic setting. 
For example we do not know of any argument which would ensure that the generator of the interacting dynamics $\tauvint_{\infty}(s,t)$ has no eigenvalues.  

If we fix a state on $\fA_{\infty}$, like for example the vacuum state for $\tau^{V}_{\infty}$, and work in its  GNS representation, replacing $C^{*}-$algebras by their weak closures, then by the same arguments as in  Subsect. \ref{sec5.2}, the dynamics $\tau^{V}_{\infty}$ and $\tauvint_{\infty}$ are implemented by unitary groups with selfadjoint generators $H_{0}$ and $H= H_{0}+ \pi_{F}(I)$. The (Hilbertian) scattering theory for the pair $H_{0}, H$ is not trivial, but nevertheless completely  understood, see e.g. \cite {A}.  In particular the Hilbertian version of asymptotic completeness was shown in \cite{A}.  However these Hilbertian results are of no use for the algebraic setting.
\end{remark}
%ici
\subsubsection{Main result II}
The following   theorem is the main result of this section.

\begin{theoreme}
 \label{6.13} The following holds: \ben
\item \[
\lim_{t\to +\infty}\omega^{V}_{0, {\rm vac}}\circ \tauvint(0,t)\circ E_{t}(A)= \omega_{\rm H, II}(A), \ A\in\gamma_{\infty}^{\rm int} \fA_{\infty}.
\] 
\item Assume moreover that $\gamma^{\rm int}_{\infty}\fA_{\infty}= \fA_{\infty}$. Then 
\[
\lim_{t\to +\infty}\omega^{V}_{0, {\rm vac}}\circ \tauvint(0,t)(A)= \omega_{\rm H, II}(A), \ A \in \fA_{J}, \ \forall J\Subset \rr.
\]
\een
\end{theoreme}

\proof Let us first prove (1). By linearity and density, it suffices to prove the theorem for  
\[
A= \A^{\rm l}\times A^{\rm r}, \ A^{\gd}= \gamma^{\rm int}_{\infty}(B^{\gd}) , \ B^{\rm l/r}\in \CAR(P^{\rm l/r}\ch_{\infty}).
\]Let us fix $\epsilon>0$. By Props. \ref{6.7} and \ref{6.12}  there exist $C_{\epsilon}, T_{\epsilon}>0$ such that
\begin{equation}
\label{ti}
\begin{array}{rl}
&\sup_{t\geq T_{\epsilon}}\| \tauvint(0,t)\circ E_{t}(A^{\rm l})- \tauvint(0, t/2+ C_{\epsilon})\circ \alpha^{t/2+ C_{\epsilon}}\circ \alpha^{- 2C_{\epsilon}}\circ\gamma^{\rm l}_{\infty}(B^{\rm l})\|\leq \epsilon,\\[2mm]
&\sup_{t\geq T_{\epsilon}}\| \tauvint(0,t)\circ E_{t}(A^{\rm r})- \tau^{V}_{0}(0,t)\circ \gamma^{\rm r}_{0}(B^{\rm r})\|\leq \epsilon.  
\end{array}
\end{equation}
We set  $\tilde{B}_{\epsilon}^{\rm l}:= \alpha^{-2C_{\epsilon}} \circ \gamma^{\rm l}_{\infty}(B^{\rm l})$ and $\tilde{B}^{\rm r}:= \gamma^{\rm r}_{0}(B^{\rm r})$. By Lemma \ref{4.6} we can by increasing $T_{\epsilon}$ ensure that
\[
\sup_{t\geq T_{\epsilon}}\| \tauvint(0,t/2+ C_{\epsilon})\circ \alpha^{t/2+ C_{\epsilon}}(\tilde{B}_{\epsilon}^{\rm l})-  \tau^{V}(0, t/2+ C_{\epsilon})\circ \alpha^{t/2+ C_{\epsilon}}(\tilde{B}^{\rm l})\| \leq \epsilon.
\]
Summarizing we have:
\begin{equation}
\label{tu}
\sup_{t\geq T_{\epsilon}}\| \tauvint(0,t)\circ E_{t}(A)- \tauv(0, t/2+ C_{\epsilon})\circ \alpha^{t/2+ C_{\epsilon}}(\tilde{B}_{\epsilon}^{\rm l})\times \tau^{V}_{0}(0,t)(\tilde{B}^{\rm r}) \|\leq C\epsilon. 
\end{equation}
We  now argue as in the proof of Lemma \ref{5.1} to obtain that:
\begin{equation}
\label{to}
\lim_{s\to +\infty} \omega^{V}_{0,{\rm vac}}\left(\tau^{V}(0,s)\circ \alpha^{s}(\tilde{B}_{\epsilon}^{\rm l})\times \tau^{V}_{0}(0,s)(\tilde{B}^{\rm r})\right)= \omega^{0}_{\infty, \beta}\widehat{\otimes}\,\, \omega^{V}_{0, {\rm vac}}(\tilde{B}_{\epsilon}^{\rm l}\times \tilde{B}^{\rm r}).
\end{equation}
To prove (\ref{to}) we use that  $\omega^{V}_{0, {\rm vac}}$ is quasi-free, and the dynamics in (\ref{to}) are free. The  cross terms  of the form:
\[
(u^{V}(0,s)f_{1}^{s}| \one_{\rr^{+}}(b^{V}_{0})u^{V}_{0}(0,t)f_{2})\hbox{ for }f_{1}\in \ch_{0}^{\rm l}, \ f_{2}\in \ch_{0}^{\rm r},
\]
vanish when $s\to +\infty$.  This is easy to see, since modulo errors which are $o(s^{0})$ in norm, the vector $u^{V}(0,s)f_{1}^{s}$ is supported in $\{|x|\leq c_{0}\}$ for $s$ large enough, while the vector $u^{V}_{0}(0,s)f_{2}$ is supported in $\{x\geq c_{1}s\}$. 

The rest of the proof is as in Lemma \ref{5.1}, using also that $\omega^{V}_{0, {\rm vac}}$ is invariant under $\tau^{V}_{0}$. 
A further observation is that the state $\omega^{0}_{\infty, \beta}$ is invariant under space translations.   Since  $\tilde{B}_{\epsilon}^{\rm l}= \alpha^{-2 C_{\epsilon}}\circ \gamma^{\rm l}_{\infty}(B^{\rm l})$, this implies that
\[
\begin{array}{rl}
 &\omega^{0}_{\infty, \beta}\widehat{\otimes}\,\, \omega^{V}_{0, {\rm vac}}(\tilde{B}_{\epsilon}^{\rm l}\times \tilde{B}^{\rm r})
 =\omega^{0}_{\infty, \beta}\widehat{\otimes}\,\, \omega^{V}_{0, {\rm vac}}(\gamma^{\rm l}_{\infty}(B^{\rm l})
 \times \gamma^{\rm r}_{0}(B^{\rm r}))\\[2mm]
 =&\omega^{\rm l}_{\infty, \beta}\widehat{\otimes}\,\,\omega^{\rm r}_{\infty, {\rm vac}}(B^{\rm l}\times B^{\rm r}).
\end{array}
\]
Therefore we can rewrite (\ref{to}) as
\[
\lim_{s\to +\infty} \omega^{V}_{0,{\rm vac}}\left(\tau^{V}(0,s)\circ \alpha^{s}(\tilde{B}_{\epsilon}^{\rm l})\times \tau^{V}_{0}(0,s)(\tilde{B}^{\rm r})\right)=\omega^{\rm l}_{\infty, \beta}\widehat{\otimes}\,\,\omega^{\rm r}_{\infty, {\rm vac}}(B^{\rm l}\times B^{\rm r}).
\]
Using also (\ref{tu}) this completes the proof of (1). Statement (2) follows from (1), since if $A\in \fA_{J}$ for some $J\Subset \rr$ then $A= E_{t}(A)$ for $t$ large enough.  \qed

\subsection{Change of initial state}\label{sec6.6}
As in Subsect. \ref{sec5.1} one can try to replace the initial state $\omega^{V}_{0, {\rm vac}}$ by another 
(even) state $\tilde{\omega}$ which belongs to the folium of $\omega^{V}_{0, {\rm vac}}$.  

There is however a difference with the situation considered in Sect. \ref{sec5}:  in Sect. \ref{sec5}  we have to consider the evolution of a right-going observable $A^{\rm r}\in \fA_{0}^{\rm r}$ under  $\tauvint(0,t)\circ \alpha^{t}$ when $t\to +\infty$: this  converges to the  limit observable $\gamma^{\rm r, int}(A^{\rm r})$, which implies that Thm. \ref{5.0} extends to any such state $\tilde{\omega}$, see Corollary \ref{5.2}.

In the present situation, we have to consider the evolution  of an  observable $B^{\rm r}\in \CAR(\ch_{0})$ under $\tau^{V}_{0}(0,t)$ (note that all observables in $\CAR(\ch_{0})$ are right-going, since $\ch_{0}= L^{2}(]z(0), +\infty[, \cc^{2})$). 
This has obviously no limit  in $\CAR(\ch_{0})$. We need to restrict ourselves to initial states $\tilde{\omega}$ on $\fA_{0}$  which have the property that
\begin{equation}
\label{bogda}
\wlim_{t\to\infty}\tilde{\omega}\circ \tau^{V}_{0}(0,t)\hbox{ exists}.
\end{equation}
Exemples of such states are states which are {\em invariant} under  the stationary interacting dynamics $\tau^{V, {\rm int}}_{0}$, which were  considered in Subsect. \ref{sec5.1}.

Let us now explain this in more details.
We first recall some facts about the algebraic scattering theory   in $\CAR(\ch_{0})$. We will use the notation introduced in Subsect. \ref{sec5.2}. It is easy to prove that the limit:
\[
\gamma_{0}^{\rm int}:= \slim_{t\to +\infty} \tau^{V,{\rm int}}_{0}(t,0)\circ \tau^{V}_{0}(0,t)
\]
exists on $\CAR(\ch_{0})$.  From Prop. \ref{6.3c} and the chain rule for wave homomorphisms, we obtain the existence of the limit
\[
\gamma^{\rm r, int}_{0}:=  \gamma_{0}^{\rm int}\circ \gamma_{0}^{\rm r}=\slim_{t\to +\infty}\tau^{V, {\rm int}}_{0}(t, 0)\circ \tau^{V}_{\infty}(0,t)\hbox{ on }\CAR(P^{\rm r}\ch_{\infty}).
\]

We obtain the following analog of Corollary \ref{5.2}.

\begin{corollary}
 \label{6.14}
 Let $\tilde{\omega}$ be an even state on $\fA_{0}$, which belongs to the folium of $\omega^{V}_{0, {\rm vac}}$ and  is invariant under $\tau^{V, {\rm int}}_{0}$. Let:
 \[
\tilde{\omega}_{\rm H, II}= (\omega^{\rm l}_{\infty, \beta}\widehat{\otimes}\,\, (\tilde{\omega}\circ \gamma^{\rm r,  int}_{0}))\circ (\gamma^{\rm int}_{\infty})^{-1}.
\]
Then:
 \ben \item
 \[
\lim_{t\to +\infty} \tilde{\omega}\circ \tau^{V, {\rm int}}(0,t)\circ E_{t}(A)= \tilde{\omega}_{\rm H, II}(A), \ A\in \gamma^{\rm int}_{\infty}\fA_{\infty}.
\]
\item Assume moreover that $\gamma^{\rm int}_{\infty}\fA_{\infty}= \fA_{\infty}$. Then 
\[
\lim_{t\to +\infty}\tilde{\omega}\circ \tauvint(0,t)(A)= \tilde{\omega}_{\rm H, II}(A), \ A \in \fA_{J}, \ \forall J\Subset \rr.
\]
\een

\end{corollary}
Note that we proved in Subsect. \ref{sec5.2} that such states $\tilde{\omega}$ exist, at least for small interactions.

\proof 
We will only sketch the proof, since it is an easy combination of the arguments in Thm. \ref{6.13} and Corollary \ref{5.2}. From (\ref{tu}) we see that modulo an error of size $\epsilon$, uniformly for $t\geq T_{\epsilon}$, we have to compute
\[
\lim_{s\to +\infty}\tilde{\omega}( \tau^{V}(0,s)\circ \alpha^{s}(\tilde{B}^{\rm l}_{\epsilon})\times \tau^{V}_{0}(0,s)(\tilde{B}^{\rm r})).
\]
We set $\tilde{B}^{\rm l}(s):=  \tau^{V}(0,s)\circ \alpha^{s}(\tilde{B}^{\rm l}_{\epsilon})$ and $\tilde{B}^{\rm r}(s):= \tau^{V}_{0}(0,s)(\tilde{B}^{\rm r})$ to simplify notation.
Since $\tilde{\omega}$ belongs to the folium of $\omega^{V}_{0, {\rm vac}}$, we can find $P=P(\psi^{*}, \psi)\in \CAR_{\rm alg}(\ch_{0})$ even,  such that:
\[
|\tilde{\omega}(B)- \omega^{V}_{0}(P^{*}BP)|\leq \epsilon \| B\|, \ B\in \CAR(\ch_{0}).
\]
By the same argument as in the proof of Corollary \ref{5.2} we have
\[
\omega^{V}_{0}(P^{*} \tilde{B}^{\rm l}(s)\tilde{B}^{\rm r}(s)P)= \omega^{V}_{0}(\tilde{B}^{\rm l}(s)P^{*}\tilde{B}^{\rm r}(s)P)+ o(s^{0}).
\]
Again the cross terms vanish when $s\to+\infty$, the terms coming from $\tilde{B}^{\rm l}(s)$ give in the limit $s\to +\infty$ the contribution $\omega^{0}_{\infty}(\tilde{B}^{\rm l})$ equal to $\omega^{\rm l}_{\infty, \beta}(B^{\rm l})$. 

The terms coming from $\tilde{B}^{\rm r}(s)$ give modulo an error of size $\epsilon$ the contribution $\tilde{\omega}(\tau^{V}_{0}(0,s)(\tilde{B}^{\rm r}))$. Now we use the hypothesis that $\tilde{\omega}$ is invariant under $\tau^{V, {\rm int}}_{0}$ hence:
\[
\begin{array}{rl}
&\tilde{\omega}(\tau^{V}_{0}(0,s)(\tilde{B}^{\rm r}))= \tilde{\omega}(\tau^{V,{\rm int}}_{0}(s,0)\circ\tau^{V}_{0}(0,s)(\tilde{B}^{\rm r}))\\[2mm]
=& \tilde{\omega}(\gamma^{\rm int}_{0}(\tilde{B}^{\rm r}))+ o(s^{0})= \tilde{\omega}(\gamma^{\rm r, int}_{0}(B^{\rm r}))+ o(s^{0}).
\end{array}
\]
We can now complete the proof as in Thm. \ref{6.13}. \qed

\section{Hawking effect III}\label{sec7}\init
In this section we study the Hawking effect in case III (see Subsect. \ref{sec1.3}).  As explained in the introduction, the interaction should now be  localized in a region
\[
\{(x,t): z(t)<x<z(t)+C, \ T-1\leq t \leq T\},
\]
and we will apply the interacting evolution to an observable $\alpha^{-z(T)}(A)$ for $A\in \fA_{0}$, letting eventually $T\to +\infty$.
Let us now make this more precise.
\subsection{Definition of the interacting dynamics}\label{sec7.1}
We fix $I\in \CAR_{0}(\ch_{0})$  as in (\ref{e4.1}) and set
\[
I(t):= \alpha^{t}(I), \ I_{T}(t):= \one_{[T-1, T]}(t)I(t),
\]
\def\ttau{\tilde{\tau}}
where  the group $\alpha^{s}$ of space translations is defined in (\ref{def-de-alpha}), and $T\gg 1$ is a parameter which will eventually tend to $+\infty$.
To be sure that $I(t)\in \fA_{t}$ we assume in this section that $z(t)\leq -t$ for all $t\geq 0$, which is not a restriction.
\begin{definition}
We denote by  $\ttau^{V, {\rm int}}_{T}(s,t)$ the  interacting dynamics (depending on the parameter $T$), constructed using Prop. \ref{app2.2}, with free dynamics $\tau^{V}(s,t)$ and time dependent interaction $\rr\ni t\mapsto I_{T}(t)$. 
\end{definition}

Our goal in this section is to study the limit:
\begin{equation}
\label{eh.2}
\lim_{T\to +\infty}\omega^{V}_{0, {\rm vac}}(\ttau_{T}^{V, {\rm int}}(0,T)\circ \alpha^{T}(A)), \ A\in \fA_{0}.
\end{equation}
\def\INT{{\rm int}}
Since $I_{T}(t)$ vanishes for $0\leq t\leq T-1$  we have:
\[
\ttau_{T}^{V, {\rm int}}(0,T-1)= \tau^{V}(0,T-1),
\]
hence
\beq\label{maino}
 \ttau^{V,\INT}_T(0,T)\circ\alpha^{T}=
\tau^{V}(0,T-1)\circ\alpha^{T-1}\circ\alpha^{-(T-1)}\circ
\ttau_{T}^{V,\INT}(T-1,T)\circ\alpha^{T}.
\eeq
Applying Thm. \ref{2.7}, the existence of the limit (\ref{eh.2}) will follow from the existence of 
\begin{equation}\label{slim_hawking3}
 \underset{T\rightarrow+\infty}{\slim}\alpha^{-(T-1)}\circ\ttau^{V,\INT}_T(T-1,
T)\circ\alpha^{T}\hbox{ on }\fA_{0}.
\end{equation}
Recall that 
$$\gamma^tf(x):=f(x+t),\;x,t\in\rr,\;f\in\ch.$$
Let us introduce the following notation: 
\begin{equation}
\label{eh.3}
\begin{array}{rl}
\hat{u}^V_T(s,t):=&\gamma^{-(T+s)}\circ u^{V}(T+s,T+t)\circ  \gamma^{T+t}\in \mathcal{U}(\ch_{0}, \ch_{0}),\\[2mm]
\hat{\tau}^V_T(s,t):=&\alpha^{-(T+s)}\circ \tau^{V}(T+s, T+t)\circ \alpha^{T+t},\ \fA_{0}\to \fA_{0},\\[2mm]
\hat{\tau}^{V, {\rm int}}_T(s,t):=&\alpha^{-(T+s)}\circ \ttau^{V,{\rm int}}_{T}(T+s, T+t)\circ \alpha^{T+t},\ \fA_{0}\to \fA_{0},
\end{array}
\end{equation}
so that the automorphism appearing in (\ref{slim_hawking3}) equals $\hat{\tau}^{V, {\rm int}}_T(-1,0)$. Note that 
\begin{equation}
\label{eh.00}
\hat{\tau}^{V}_{T}(s,t)(\psi^{(*)}(f))= \psi^{(*)}(\hat{u}_{T}^{V}(s,t)f), \ f\in \ch_{0},
\end{equation}
and that $\{\hat{u}^V_T(s,t)\}_{s,t\in \rr}$ is a two-parameter propagator, while $\{\hat{\tau}^V_T(s,t)\}_{s,t\in \rr}$ and $\{\hat{\tau}^{V,{\rm int}}_T(s,t)\}_{s,t\in \rr}$ are two-parameter quantum dynamics.
\subsection{Preparations}\label{sec7.2}
We start by  considering the  limit (\ref{slim_hawking3}) for $I=0$.
\begin{lemma}The strong limit
 \begin{equation}\nonumber
  \hat{u}^0_\infty(s,t):=\slim_{T\rightarrow+\infty}\hat{u}^0_T(s,t)
 \end{equation}
exists on $\ch_{0}$ and the convergence is uniform in   $a\leq s \leq t\leq b$ for any $a\leq b$. Moreover $\{\hat{u}^0_\infty(s,t)\}_{s,t\in \rr}$ is a two-parameter propagator  given by
$$\hat{u}^0_{\infty}(s,t)f=
\begin{pmatrix}
\gamma^{2(t-s)}f_1
\\
f_2
\end{pmatrix}, \ f\in \ch_{0}.
$$
\end{lemma}
\begin{remark}
The convergence above holds a priori only for $s\leq t$. Nevertheless the limit $\hat{u}^{0}_{\infty}(s,t)$ is defined for all $s,t\in \rr$.
\end{remark}
\proof
It is easy to obtain an explicit expression for $\hat{u}^{0}_{T}(s,t)$. In fact from \cite[Lemme VI.3]{Ba1} we know that if $\psi(s, \cdot)= u^{0}(s,t)f(\cdot)$ for $f\in \ch_{t}$ then:
\[
\begin{array}{rl}\psi_1(s,x)=&f_1(x-s+t)\\[2mm]
\psi_2(s,x)=& \left\{
\begin{array}{l}\lambda\circ\tau(x+s)^{-1}f_1(x+t+s-2\tau(x+s)) \hbox{ for } z(s)<x<z(t)+t-s,
\\[2mm]
f_2(x-t+s) \hbox{ for } x>z(t)+t-s,
\end{array}
\right.
\end{array}
\]
where the reflection coefficient $\lambda$ is defined in  Subsect. \ref{sec1.2}, and the function $y\mapsto \tau(y)$ is  the inverse of the function $s\mapsto s+z(s)$ (see \cite[Equ. VI.40]{Ba1}).

\def\tz{\tilde{z}}
From this a routine computation gives that:
\[
\begin{array}{rl}(\hat{u}^{0}_{T}(s,t)f)_{1}(x)=& f_{1}(x+2(t-s)),\\[2mm]
(\hat{u}^{0}_{T}(s,t)f)_{2}(x)=& \left\{
\begin{array}{l}\lambda\circ\tau(x)^{-1}f_1(x+2(T+t)-2\tau(x)), \\[2mm]
\hbox{ for } \tz(T+s)<x<\tz(T+t),
\\[2mm]
f_2(x), \hbox{ for } x>\tz(T+t),
\end{array}
\right.
\end{array}
\]
where $\tz(\sigma):= \sigma+z(\sigma)\in o(\sigma^{0})$, by (\ref{definitionsurface}).
Using this fact and that $f$ is compactly supported, we easily see that 
$$\lim_{T\rightarrow+\infty}\hat{u}^0_T(s,t)f=
\begin{pmatrix}
\gamma^{2(t-s)}f_1
\\
f_2
\end{pmatrix},
$$
uniformly for $a\leq s\leq t\leq b$.  This completes the proof. \qed

We now establish the same result for arbitrary $V$.
\begin{lemma}
The limit
 \begin{equation}\nonumber
  \slim_{T\rightarrow+\infty}\hat{u}^V_T(s,t)=\hat{u}^0_\infty(s,t),
 \end{equation}
exists, and the convergence is uniform in   $a\leq s \leq t\leq b$ for any $a\leq b$. 
\end{lemma}
\proof
From Duhamel's formula we obtain that
\[
\|u^{V}(T+s, T+t)f- u^{0}(T+s, T+t)f\|\leq \int_{s}^{t}\|V u^{0}(T+\sigma, T+t)f\|d\sigma,
\]
hence 
\[
\|\hat{u}^{V}_{T}(s,t)f- \hat{u}^{0}_{T}(s,t)f\|\leq \int_{s}^{t}\| V u^{0}(T+\sigma, T+t)\gamma^{T+t}f\| d\sigma.
\]
By the usual density argument we can assume that $\supp f\subset[0, b]$.  We deduce from this that  $\supp u^{0}(T+\sigma, T+t) \gamma^{T+t}f\subset [-T+ o(T^{0}), b-T+ o(T^{0})]$, uniformly for $a\leq s\leq \sigma\leq t\leq b$.  Using the decay property of $V$ near $-\infty$, this implies that
\[
\lim_{T\to +\infty}\sup_{-1\leq s \leq t\leq 0}\|\hat{u}^{V}_{T}(s,t)f- \hat{u}^{0}_{T}(s,t)f\|=0,
\]
which completes the proof of the lemma. \qed

By (\ref{eh.00})  we obtain 
\begin{proposition}\label{h1}
The strong limit 
\[
\hat{\tau}^{0}_{\infty}(s,t):= \slim_{T\to +\infty} \hat{\tau}^{V}_{T}(s,t)
\]
exists on $\fA_{0}$ and the convergence is uniform in   $a\leq s \leq t\leq b$ for any $a\leq b$.
\end{proposition}
\subsection{Hawking effect III}\label{sec7.3}
\begin{proposition}\label{h2}
Let $\hat{\tau}^{0,{\rm int}}_{\infty}(s,t)$ be the interacting dynamics obtained from Prop. \ref{app2.2} from the free dynamics $\hat{\tau}^{0}_{\infty}(s,t)$ and the interaction $I$.
Then 
\[
 \slim_{T\to +\infty}\hat{\tau}^{V,{\rm int}}_{T}(s,t)=\hat{\tau}^{0,{\rm int}}_{\infty}(s,t),
\]
and the convergence is uniform in   $a\leq s \leq t\leq b$. 
\end{proposition}
\proof 
Let $R_{T}(s,t)$ the unitary operator in Prop. \ref{app2.2} for the time-dependent interaction $I_{T}(\cdot)$.  Then from  (\ref{eh.3}), we see that:
\[
\begin{array}{rl}
&\hat{\tau}_{T}^{V,{\rm int}}(s,t)(A)\\[2mm]
=& \alpha^{-(T+s)}R_{T+s}(T+s, T+t)\times \hat{\tau}^{V}_{T}(s,t)(A)\times \alpha^{-(T+s)} R_{T+s}(T+s, T+t)^{*}, \ A\in \fA_{0}.
\end{array}
\]
Using Prop. \ref{h1} it hence suffices to show that
\[
R_{\infty}(s,t):=\lim_{T\to +\infty}\alpha^{-(T+s)}R_{T+s}(T+s, T+t)\hbox{ exists}.
\]
By Lemma \ref{app2.1}, we have:
\beq\label{eh.4}
\begin{array}{rl}
&\alpha^{-(T+s)}R_{T}(T+s, T+t)\\[2mm]
=& \sum_{n\geq 0}(-\i)^{n}\int_{T+s\leq t_{n}\leq \cdots \leq t_{1}\leq T+t}\alpha^{-(T+s)}I_{T}(T+s, t_{n})\cdots \alpha^{-(T+s)}I_{T}(T+s, t_{1})dt_{n}\cdots dt_{1}\\[2mm]
=& \sum_{n\geq 0}(-\i)^{n}\int_{s\leq t_{n}\leq \cdots \leq t_{1}\leq t}\alpha^{-(T+s)}I_{T}(T+s, T+t_{n})\cdots \alpha^{-(T+s)}I_{T}(T+s,T+ t_{1})dt_{n}\cdots dt_{1}\\[2mm]
\end{array}
\eeq
Note that:
\[
\alpha^{-(T+s)}I_{T}(T+s, \sigma+T)=\one_{[-1, 0]}(s)\hat{\tau}_{T}^{V}(s, \sigma)(I).
\]
By Prop. \ref{h1} we know that
\[
\lim_{T\to +\infty} \hat{\tau}_{T}^{V}(s, \sigma)(I)= \hat{\tau}_{\infty}^{0}(s, \sigma)(I),
\]
uniformly for $-1\leq s\leq \sigma\leq 0$. Therefore using that the convergence of the series in (\ref{eh.4}) is uniform in $T$, we can pass to the limit under the sum and integrals, which range over compact regions.  The limit
\[
\hat{R}_{\infty}(s,t)= \lim_{T\to +\infty}\alpha^{-(T+s)}R_{T}(T+s, T+t)
\]
equals the unitary operator obtained in Prop. \ref{app2.2} from the free dynamics $\hat{\tau}^{0}_{\infty}(s, t)$ and the interaction $I$.   Therefore
\[
\lim_{T\to \infty}\hat{\tau}^{V,{\rm int}}_{T}(T+s, T+t)(A)= \hat{R}_{\infty}(s,t)\times \hat{\tau}^{0}_{\infty}(s,t)(A)\times\hat{R}_{\infty}(s,t)^{*}= \hat{\tau}_{\infty}^{0, {\rm int}}(s,t)(A).
\]
This completes the proof of the proposition. \qed

We can now formulate the main result of this section.  We first define the limiting state. Let $\gamma^{\rm r}=\slim_{t\to +\infty}\tau^{V}(0,t)\circ \alpha^{t}$ on $\fA^{\rm r}_{0}$, obtained as in Prop. \ref{4.7} if the interaction $I$ vanishes. Note that $\gamma^{\rm r}$ is implemented by the classical wave operator $\slim_{t\to+\infty} u^{V}(0,t)\circ \gamma^{t}$ on $\ch_{0}^{\rm r}$.

Let
\[
\omega^{\rm free}_{\rm H}:=  \omega^{0}_{\infty, \beta}\hat{\otimes}(\omega^{V}_{0,{\rm vac}}\circ \gamma^{\rm r}).
\]
This is the limiting state obtained by Bachelot in \cite{Ba1} in the case when the interaction vanishes.
\begin{theoreme}\label{h5}
\[
\lim_{T\to \infty} \omega^{V}_{0, {\rm vac}}\circ \ttau_{T}^{V,{\rm int}}(0, T)\circ \alpha^{T}(A)= \omega_{\rm H}^{\rm free }\circ \hat{\tau}^{0, {\rm int}}_{\infty}( -1,0)(A), \ A\in \fA_{0}.
\]
\end{theoreme}
\proof The result follows from Prop. \ref{h2}, formula (\ref{maino}) and   Thm.  \ref{5.0} in the case  $I=0$, (which in this case was already obtained  in \cite{Ba1}). 
\qed

 A similar result can be obtained if we replace the initial state $\omega^{V}_{0, {\rm vac}}$ by  another state $\tilde{\omega}$ as in Corollary \ref{5.2}.
\appendix
\section{}\label{secapp1}\init
In this appendix we recall various well-known facts about $\CAR$ algebras, fermionic Fock spaces and groups of $*-$isomorphisms on $C^{*}-$algebras.
\subsection{$\CAR$ algebras}\label{secapp1.1}
\subsubsection{$\CAR$ algebras}
 Let $\ch$ be a (complex) Hilbert space, with scalar product denoted by $(\cdot | \cdot)$.
\begin{definition}\label{app1.1}
The {\em  algebraic CAR algebra over} $\ch$, denoted by $\CAR_{\rm alg}(\ch)$  is the  unital $*-$algebra generated by the generators 
$\one$, $\psi(h)$, $h\in  \ch$ and the relations:
 \[
\begin{array}{rl}
&\psi(h_{1}+ h_{2})= \psi(h_{1})+ \psi(h_{2}), \ h_{i}\in \ch,\\[2mm]
&\psi(\lambda h)= \overline{\lambda}\psi(h), \ h\in \ch, \ \lambda\in \cc,\\[2mm]
&[\psi(h_{1}), \psi(h_{2})]_{+}= 0, \ [\psi(h_{1}), \psi^{*}(h_{2})]_{+}= (h_{1}| h_{2})\one.
\end{array}
\]
\end{definition}
It is well known that $\CAR_{\rm alg}(\ch)$ is simple, hence has a unique $C^{*}-$norm. A concrete expression for this norm can be  obtained by taking the representation $\pi:\  \CAR_{\rm alg}(\ch)\to B(\Gamma_{\a}(\ch))$ with $\pi(\psi^{(*)}(h)):= a^{(*)}(h)$ for $h\in \ch$, see Subsect. \ref{secapp.3bis} below. (This corresponds to the choice of $\ii= \i$ as K\"{a}hler structure).
\begin{definition}\label{app1.2}
 The {\em CAR $C^{*}-$algebra over }$\ch$, denoted by $\CAR(\ch)$ is the completion of $\CAR_{\rm alg}(\ch)$ for its unique $C^{*}-$norm.
\end{definition}
An element of $\CAR_{\rm alg}(\ch)$ can be written in a unique way in {\em normal ordered form}, i.e. with $\psi^{*}$'s to the left of $\psi$'s. This allows to define unambiguously   the {\em monomials}. A monomial $A$ has a degree denoted by $\deg A$. Sometimes we will also use the {\em bi-degree} $(n,p)$, where $n$ is the number of factors of $\psi^{*}$, $p$ the number of factors of $\psi$.

\subsubsection{Parity}
$\CAR(\ch)$ is equipped with the {\em parity automorphism} $P$, defined by 
\[
P \psi^{(*)}(h):= \psi^{(*)}(-h), \ h\in \ch.
\] 

We denote by $\CAR_{0}(\ch)$, resp. $\CAR_{1}(\ch)$ the subspace of even, resp. odd elements of $\CAR(\ch)$. $\CAR_{0}(\ch)$ is a $C^{*}-$sub-algebra of $\CAR(\ch)$. 

A state $\omega$ on $\CAR(\ch)$ is {\em even} if $\omega\circ P=\omega$, or equivalently $\omega=0$ on $\CAR_{1}(\ch)$.

A $*-$automorphism $\alpha$ of $\CAR(\ch)$ is {\em even} if $ P\circ \alpha= \alpha\circ P$.
 \subsubsection{Conditional expectations}\label{astolito}
 If $\ch_{1}$  is  a closed subspace of $\ch$, then  $\CAR(\ch_{1})$  is a $C^{*}-$subalgebra of $\CAR(\ch)$. The converse  construction is as follows:  define $E_{\ch_{1}}: \CAR_{\rm alg}(\ch)\to \CAR_{\rm alg}(\ch_{1})$ by 
\[
E_{u} \psi^{(*)}(f):= \psi^{(*)}(\pi f), \ f\in \ch,
\] 
where $\pi: \ch\to \ch_{1}$ is the orthogonal projection.
Then $E_{\ch_{1}}$ extends  as a $*-$homomorphism from $\CAR(\ch)$ to $\CAR(\ch_{1})$.  This can be easily checked by using the Fock representations of $\CAR(\ch)$ (resp. $\CAR(\ch_{1})$) on $\Gamma_{\rm a}(\ch)$ (resp. $\Gamma_{\rm a} (\ch_{1})$) and the second quantized map  $\Gamma(\pi)$. Moreover  
if $\{\ch_{i}\}_{i\in I}$ is an increasing net of closed subspaces of $\ch$ with $\cup_{i\in I}\ch_{i}$ dense in $\ch$ then 
\[
\slim_{i}E_{\ch_{i}}= \one, \hbox{ in }\CAR(\ch).
\]
\subsection{Fermionic exponential law}\label{secapp1.2}
\subsubsection{$\zz_{2}-$graded tensor product}
 Let $\ch_{i}$, $i=1,2$ be two Hilbert spaces.  We equip the vector space $\CAR_{\alg}(\ch_{1})\otimes \CAR_{\alg}(\ch_{2})$ with the  $*-$algebra structure defined by:
 \beq\label{eapp.1}
\begin{array}{rl}
(a_{1}\otimes a_{2})\cdot (b_{1}\otimes b_{2}):=& (-1)^{\deg a_{2}\deg b_{1}} a_{1}b_{1}\otimes a_{2}b_{2},\\[2mm]
(a_{1}\otimes a_{2})^{*}:=& (-1)^{ \deg a_{1}\deg a_{2}} a_{1}^{*}\otimes a_{2}^{*},
\end{array}
\eeq
for $a_{i}$, $b_{i}$ monomials in $\CAR_{\rm alg}(\ch_{i})$, and extended to $\CAR_{\alg}(\ch_{1})\otimes \CAR_{\alg}(\ch_{2})$ by linearity.  The resulting $*-$algebra is denoted by $\CAR_{\alg}(\ch_{1})\widehat{\otimes}\,\,\CAR_{\alg}(\ch_{2})$.
\begin{lemma}\label{app1.3}
 The map 
 \[
\begin{array}{rl}
I: \CAR_{\alg}(\ch_{1}\oplus \ch_{2})\to \CAR_{\alg}(\ch_{1})\widehat{\otimes}\,\,\CAR_{\alg}(\ch_{2})\\[2mm]
\psi^{(*)}(h_{1}\oplus h_{2})\mapsto \psi^{(*)}(h_{1})\otimes \one + \one \otimes \psi^{(*)}(h_{2})
\end{array}
\]
extends as a $*-$isomorphism.
\end{lemma}
\begin{remark}\label{rem1}
If $u: \ch\to \tilde{\ch}$ is an isometry then the map $\psi^{(*)}(h)\mapsto \psi^{*}(uh)$ extends to an  $*-$homomorphism from $\CAR(\ch)$ to $\CAR(\tilde{\ch})$.This allows to see $\CAR(\ch_{i})$ $i=1,2$ as $*-$sub-algebras of $\CAR(\ch_{1}\oplus \ch_{2})$. If $A_{i}\in \CAR(\ch_{i})$ let us still denote by $A_{i}$ its image in $\CAR(\ch_{1}\oplus \ch_{2})$. Then clearly 
we have 
\[
I(A_{1}A_{2})= A_{1}\otimes A_{2}, \hbox{ for }A_{i}\in \CAR(\ch_{i}).
\]

\end{remark}
\begin{definition}\label{app1.4}
The {\em $\zz_{2}-$graded tensor product} $\CAR(\ch_{1})\widehat{\otimes}\,\,\CAR(\ch_{2})$ is the completion of $\CAR_{\alg}(\ch_{1})\widehat{\otimes}\,\CAR_{\alg}(\ch_{2})$ for the norm $\| I^{-1}\cdot \|$.  
\end{definition}
\subsubsection{Tensor product of states and automorphisms}
\begin{lemma}\label{app1.5}
 Let $\omega_{i}$ be a state on $\CAR(\ch_{i})$, $i=1,2$.  Assume that $\omega_{1}$ is {\em even}, i.e. $\omega_{1}\circ P_{1}= \omega_{1}$. Then $\omega_{1}\otimes \omega_{2}$ is a state on $\CAR(\ch_{1})\widehat{\otimes}\,\CAR(\ch_{2})$.
\end{lemma}
\proof It suffices to check positivity. If $A= \sum_{1}^{n}\lambda_{i}a_{1i }\otimes a_{2i}$, where $a_{ki}$ are monomials, $\lambda_{i}\in \cc$, then from (\ref{eapp.1}) we get that:
\[
A^{*}A= \sum_{i,j}\overline{\lambda}_{i}\lambda_{j} (-1)^{ d_{2i}(d_{1i}+ d_{1j})} a_{1i}^{*}a_{1j}\otimes a_{2i}^{*}a_{2j},
\]
for $d_{ki}= \deg a_{ki}$. Since $\omega_{1}$ is even, we obtain that:
\[
\omega(A^{*}A)= \sum_{i,j}\overline{\lambda}_{i}\lambda_{j} \omega_{1}(a_{1i}^{*}a_{1j}) \omega_{2}(a_{2i}^{*}a_{2j}).
\]
The positivity follows from the well-known fact that the pointwise product of two positive selfadjoint matrices is positive selfadjoint. \qed

\begin{definition}\label{app1.6}
 Let $\omega_{i}$ be a state on  $\CAR(\ch_{i})$, $i=1,2$ with $\omega_{1}$ even. The {\em $\zz_{2}-$graded tensor product} 
 $\omega_{1}\widehat{\otimes}\,\omega_{2}$ is the state on $\CAR(\ch_{1}\oplus \ch_{2})$ equal to
 $\omega_{1}\otimes \omega_{2}\circ I$.
 \end{definition}
Similarly one easily check that if $\alpha_{i}$, $i=1,2$ are {\em even} $*-$automorphisms of $\CAR(\ch_{i})$ then $\alpha_{1}\otimes \alpha_{2}$ is a $*-$automorphism of $\CAR(\ch_{1})\widehat{\otimes}\,\CAR(\ch_{2})$.

\begin{definition}\label{app1.7}
 Let $\alpha_{i}$ $i=1,2$ be even $*-$automorphisms of $\CAR(\ch_{i})$. The {\em $\zz_{2}-$graded tensor product}  $\alpha_{1}\widehat{\otimes}\,\alpha_{2}$ is the $*-$automorphism of $\CAR(\ch_{1}\oplus \ch_{2})$ equal to $I^{-1}\circ (\alpha_{1}\otimes \alpha_{2})\circ I$.
\end{definition}

\subsection{Quasi-free states}\label{secapp1.3}
We now recall some well-known facts on quasi-free states.
\begin{definition}\label{def-quasi}
 A state $\omega$ on $\CAR(\ch)$is a (gauge-invariant) {\em quasi-free state} if
 \[
\begin{array}{rl}
&\omega(\prod_{i=1}^{n}\psi^{*}(f_{i})\prod_{i=1}^{p}\psi(g_{i}))=0, \hbox{ for }n\neq p,\\[2mm]
&\omega(\prod_{i=1}^{n}\psi^{*}(f_{i})\prod_{i=1}^{n}\psi(g_{i}))=\sum_{\sigma\in S_{n}}\epsilon(\sigma) \prod_{i=1}^{n}\omega(\psi^{*}(f_{i})\psi(g_{\sigma(i)})).
\end{array}
\]
The bounded selfadjoint operator $c$ on $\ch$ defined by
\[
\omega(\psi^{*}(f)\psi(g))=: (g| cf)_{\ch}, \ f, g\in \ch
\]
is called the {\em covariance}  of $\omega$.
\end{definition}
It is well-known that a necessary and sufficient condition for a selfadjoint operator $c$ to be the covariance of a quasi-free state is:
\begin{equation}
\label{eapp1.3}
0\leq c\leq \one.
\end{equation} 
\subsection{Fermionic Fock spaces}\label{secapp.3bis}
Let $\cZ$ be a complex Hilbert space. The {\em fermionic Fock space} over $\cZ$ is the Hilbert space
\[
\Gamma_{\rm a}(\cZ):=\mathop{\oplus}\limits_{n=0}^{\infty}\otimes_{\rm a}^{n}\cZ,
\]
where $\otimes_{\rm a}^{n}\cZ$ denotes the anti-symmetric $n-$th tensor power of $\cZ$.  On $\cZ$ one defines the {\em creation/annihilation operators} $a^{*}(u)$,  $a(u)$ (see e.g. \cite{RS}) satisfying the canonical anti-commutation relations:
\[
[a(u), a(v)]_{+}= [a^{*}(u), a^{*}(v)]_{+}=0, \  [a(u), a^{*}(v)]_{+}=(u| v)_{\cZ}\one,
\]
where $[\cdot, \cdot]_{+}$ denotes the anti-commutator. If $b$ is an operator acting on $\cZ$ ,  $\d\Gamma(b)$ is  its second quantization, acting on $\Gamma_{\rm a}(\cZ)$ (see e.g. \cite{RS}).  The unit  vector $\Omega= (1, 0,\dots)\in \Gamma_{\rm a}(\cZ)$ is called the {\em vacuum}.
\subsection{Fock representation associated to a K\"{a}hler structure}
Let $\ch$ be a complex Hilbert space.  We denote its complex structure by $\i$ and its scalar product by $(\cdot|\cdot)$. The space $\ch$ considered as a real vector space will be denoted by $\ch_{\rr}$.

A {\em K\"{a}hler structure} on $\ch$ is  a unitary anti-involution $\ii$ acting on $\ch$. Note that $\kappa:= -\i \ii$ is a selfadjoint involution. Therefore we can split $\ch$ as $\ch^{+}\oplus \ch^{-}$, where $\ch^{\pm}:= \one_{\{\pm 1\}}(\kappa)\ch$. We also set $f^{\pm}:=  \one_{\{\pm 1\}}(\kappa)f$ for $f\in \ch$.

Let us denote by $\cZ$ the real vector space $\ch_{\rr}$ equipped with the complex structure $\ii$. 
We can furthermore turn $\cZ$ into a Hilbert space by equipping it with the scalar product:
\[
 (u|v)_{\cZ}:= (u^{+}|v^{+})+ (v^{-}|u^{-}).
\]
The Hilbert space $\cZ$ is called the {\em one-particle space} (associated to the K\"{a}hler structure $\ii$).

One can then define the {\em Fock representation} of $\CAR(\ch)$ in $\Gamma_{\rm a}(\cZ)$ by setting
\[
\psi_{F}(f)= \pi_{F}(\psi(f)):= \12 a^{*}(f^{+})+ \12 a(f^{-}), \ f\in \ch.
\] 
The operator $Q:= \d\Gamma(\kappa)$ acting on $\Gamma_{\rm a}(\cZ)$ is usually called the {\em charge operator}. One has:
\beq\label{deb1}
\e^{\i \theta Q} \psi_{F}^{(*)}(f)\e^{- \i \theta Q}= \psi_{F}(\e^{\i \theta}f), \ f\in \ch, \ \theta\in \rr.
\eeq
If $b$ is a selfadjoint operator on $\ch$ which commutes with $\ii$, then $c= \kappa b$ is selfadjoint on $\cZ$. The operator  
$H= \d\Gamma(c)$ acting on $\Gamma_{\rm a}(\cZ)$ is usually called the (quantum) {\em Hamiltonian}.  Note that $H\geq 0$ and $\Omega\in \Gamma_{\rm a}(\cZ)$ is its unique ground state. One has:
\beq\label{deb2}
\e^{\ i tH} \psi_{F}(f)\e^{- \i tH}= \phi_{F}(\e^{\i tb}f), \ f\in \ch, \ t \in \rr,
\eeq
i.e. the unitary group $\e^{\i tH}$ implements the dynamics generated by $\e^{\i tb}$ in the Fock representation.
Note that to conform with the common usage, we denoted by $\i$ in (\ref{deb1}) and (\ref{deb2}) the complex structure on $\Gamma_{\rm a}(\cZ)$.

\subsection{Some auxiliary results}\label{secapp.4}
The following lemma is well known.
\begin{lemma}\label{app2.1}
 Let $\fA$ a $C^{*}-$algebra and $\rr\ni t\mapsto H(t)\in \fA$ a continuous map with $H(t)= H^{*}(t)$.
 Then there exists a unique  $C^{1}$ map 
 \[
\rr^{2}\ni (s,t)\mapsto U_{H(\cdot)}(s,t)\in \fA,
\]
such that:
\[
\begin{array}{rl}
i)&\p_{t}U_{H(\cdot)}(s,t)= -\i U_{H(\cdot)}(s,t)H(t), \ s,t \in \rr,\\[2mm]
ii)&\p_{s}U_{H(\cdot)}(s,t)= \i H(s)U_{H(\cdot)}(s,t),\ s, t\in \rr,\\[2mm]
iii)& U_{H(\cdot)}(t,t)= \one.
\end{array}
\]
Moreover one has
\[
\begin{array}{rl}
iv)& U_{H(\cdot)}(s,t)= 
\mathop{\sum}\limits _{n=0}^{\infty}(-\i)^{n}\int_{s\leq t_{n}\leq \cdots\leq t_{1}\leq t}H(t_{n})\cdots H(t_{1})dt_{n}\dots dt_{1},\\[2mm]
v)& U_{H(\cdot)}(s,t')U_{H(\cdot)}(t',t)= U_{H(\cdot)}(s,t), \ s,t',t\in \rr,\\[2mm]
vi)&U_{H(\cdot)}(s,t)\hbox{ is unitary in }\fA.
\end{array}
\]
\end{lemma}
We now further study $U_{H(\cdot)}(s,t)$ if $t\mapsto H(t)$ is obtained from a  quantum dynamics. 

It is natural  to  generalize the framework of Def. \ref{3.1}. Instead of choosing $\fA_{t}= \CAR(\ch_{t})$ we can assume that $\fA_{t}$ for $t\in\rr$ are $C^{*}-$algebras with $\fA_{t}\subset \fA_{s}\subset \fA_{\infty}$  for $s\leq t$  for some $C^{*}-$algebra $\fA_{\infty}$. Moreover we can assume that for each $t_{0}\in \rr$, there exists a $*-$sub-algebra $\tilde{\fA}_{t_{0}}$ dense in $\fA_{t_{0}}$ such that an $A\in \tilde{\fA}_{t_{0}}$ belongs to $\fA_{t}$ for $t$ near $t_{0}$.  Then the obvious generalization  of Def. \ref{3.1} makes sense.
\begin{proposition}\label{app2.2}
 Let $\fA_{t}$ for $t\in\rr$ be a family of $C^{*}-$algebras satisfying the above conditions and $\tau^{0}(s,t):\ \fA_{t}\tilde{\to} \ \fA_{s}$ a quantum dynamics. Let $\rr\ni t\mapsto I(t)\in \fA_{\infty}$ a continuous map with $I(t)= I^{*}(t)$ and $I(t)\in \fA_{t}$ for $t\in \rr$.  Set:
  \[
 I(s,t):= \tau^{0}(s,t)(I(t))\in \fA_{s}, \ R_{s}(t', t):= U_{I(s, \cdot)}(t',t)\in U(\fA_{s}).
 \]
Then
\ben
\item \[
\tau^{0}(s,t')R_{t'}(t',t)= R_{s}(t',t), \ s, t',t\in \rr;
\]
\item Set \[
\tau(s,t)(A):= R_{s}(s,t) \tau^{0}(s,t)(A) R_{s}(s,t)^{*}, \ s,t\in \rr.
\] 
Then $\tau(s,t): \ \fA_{t}\ \tilde{\to}\ \fA_{s}$ is a  quantum dynamics.  
\een
\end{proposition}
\proof Statement (1)  follows by differentiating both members w.r.t. $t$ and using the uniqueness result in Lemma \ref{app2.1}. (2) follows from (1).
 \qed

 The following remark will hopefully clarify the meaning of $\tau(s,t)$ constructed in Prop. \ref{app2.2}.
\begin{remark}
 Assume that $\fA_{t}\equiv \fA$ and let $H_{0}= H_{0}^{*}\in \fA$ and $\rr\ni t\mapsto I(t)\in \fA$ be continuous. Set  $\tau^{0}(s,t)A= \e^{\i (s-t)H_{0}}A\e^{- \i (s-t)H_{0}}$ and let $\tau(s,t)$ be obtained from Prop. \ref{app2.2}. Then 
 $\tau(s,t)A=  U(s,t)AU(t,s)$ where $\{U(s,t)\}_{s,t\in \rr}$ is the two-parameter propagator obtained from Lemma \ref{app2.1} for $H(t)= H_{0}+ I(t)$.
\end{remark}

\end{document}